\documentclass[twocolumn,aps,prr,superscriptaddress,nofootinbib]{revtex4-2}
\usepackage{graphicx}
\usepackage{dcolumn}
\usepackage{bm}
\usepackage[latin5]{inputenc}
\usepackage{amssymb,amsmath,amsthm,amsbsy,mathptmx,gensymb}
\DeclareMathAlphabet{\mathcal}{OMS}{cmsy}{m}{n}
\usepackage{tabularx}
\usepackage{xcolor}
\usepackage{array}
\newcolumntype{P}[1]{>{\centering\arraybackslash}p{#1}}
\usepackage{mathtools}
\usepackage{oubraces}
\usepackage{dsfont}
\usepackage[title, titletoc]{appendix}
\usepackage{graphicx}
\usepackage{rotating}
\usepackage[most]{tcolorbox}
\usepackage{enumerate}
\usepackage{subfloat}
\usepackage{multirow}
\usepackage{balance}
\usepackage{lastpage}
\usepackage[colorlinks=true,linkcolor=blue,citecolor=blue,urlcolor=blue]{hyperref}%
\usepackage{xcolor, soul}
\definecolor{light-gray}{gray}{0.85}
\sethlcolor{light-gray}
\usepackage{cleveref}
\Crefname{subfigures}{figure}{figures}
\Crefname{subfigures}{Figure}{Figures}
\usepackage{lipsum}

\Crefname{subfigures}{figure}{figures}
\Crefname{subfigures}{Figure}{Figures}

\newcommand{\nc}{\newcommand}
\nc{\rnc}{\renewcommand}
\nc{\beg}{\begin{equation}}
\nc{\eeq}{{\end{equation}}}
\nc{\beqa}{\begin{eqnarray}}
\nc{\eeqa}{\end{eqnarray}}
\nc{\lbar}[1]{\overline{#1}}
\nc{\bra}[1]{\langle#1|}
\nc{\ket}[1]{|#1\rangle}
\nc{\ketbra}[2]{|#1\rangle\!\langle#2|}
\nc{\braket}[2]{\langle#1|#2\rangle}

\begin{document}
\title{Classical and Quantum Orbital Correlations in the Molecular Electronic States}
\author{Onur Pusuluk}
\email{onur.pusuluk@gmail.com}
\affiliation{Department of Physics, Ko{\c{c}} University, 34450 Sar{\i}yer, Istanbul, Turkey}
\author{Mahir H. Ye\c{s}iller}
\affiliation{Department of Physics, Ko{\c{c}} University, 34450 Sar{\i}yer, Istanbul, Turkey}
\author{G\"{o}khan Torun}
\affiliation{Department of Physics, Ko{\c{c}} University, 34450 Sar{\i}yer, Istanbul, Turkey}
\author{\"{O}zg\"{u}r E. M\"{u}stecapl{\i}o\u{g}lu}
\affiliation{Department of Physics, Ko{\c{c}} University, 34450 Sar{\i}yer, Istanbul, Turkey}
\author{Ersin Yurtsever}
\affiliation{Department of Chemistry, Ko{\c{c}} University, 34450 Sar{\i}yer, Istanbul, Turkey}
\author{Vlatko Vedral}
\affiliation{Department of Physics, University of Oxford, Parks
Road, Oxford, OX1 3PU, UK} \affiliation{Centre for Quantum
Technologies, National University of Singapore, 3 Science Drive 2,
Singapore 117543, Singapore}



\begin{abstract}

The quantum superposition principle has been extensively utilized in the quantum mechanical description of the bonding phenomenon. It explains the emergence of delocalized molecular orbitals and provides a recipe for the construction of near-exact electronic wavefunctions. On the other hand, its existence in composite systems may give rise to nonclassical correlations that are regarded now as a resource in quantum technologies. Here, we approach the electronic ground states of three prototypical molecules from the point of view of fermionic information theory. For the first time in the literature, we properly decompose the pairwise orbital correlations into their classical and quantum parts in the presence of superselection rules. We observe that quantum orbital correlations can be stronger than classical orbital correlations though not often. Also, quantum orbital correlations can survive even in the absence of orbital entanglement depending on the symmetries of the constituent orbitals. Finally, we demonstrate that orbital entanglement would be underestimated if the orbital density matrices were treated as qubit states.

\end{abstract}

\maketitle


\section{Introduction}

One of the principal contributions of quantum mechanics to chemistry has been the description of chemical bonding as the quantum superposition phenomenon~\cite{1999WeinholdCbonding}. In valence bond (VB) theory~\cite{1960_Pauling}, the increase in the stability of bonded atoms originates from the superposition of alternative electronic structures known as VB or resonance structures. Molecular orbital theory~\cite{1952_Coulson} suggests a different description for the same reality. What stabilizes the interacting atoms by a certain amount of energy is the delocalization of the electrons participating in this interaction throughout the molecule. Namely, the electrons occupy delocalized orbitals made from the quantum superposition of atomic orbitals. These two quantum mechanical theories are equivalent to each other at the limit~\cite{1978_MO2VB, 2021_MO2VB}, and both explain the stability of molecules as a decrease in energy due to the principle of quantum superposition.

The quantum states that arise from the quantum superposition principle are now regarded as a resource as real as energy for new quantum technologies~\cite{2003_QTech}. Quantum coherence and correlations~\cite{2012ModiCorrelations, 2016AdessoQC} are the characteristic traits of such states that enforce the entire departure of quantum materials from classical lines of thought.

The quantum entanglement is considered to be the perfect example of quantum correlations and its role as a resource in quantum technologies is well established~\cite{2009HorodeckiEnt, 2009_EntTech, 2018_EntTech}. However, it constitutes only a subset of the most general quantum correlations known as quantum discord~\cite{Vedral-2001, Zurek-2002}.

The discord can survive in the dissipative environments that wash away all the entanglement~\cite{2010_ThermalDiscord}. Also, it can serve a useful role as a resource in some tasks that are otherwise impossible, even in the absence of the entanglement. These include the detection of quantum phase transitions~\cite{2008RaoulQDResource}, the remote state preparation for quantum information processing~\cite{2012DakicRemoteSP}, the secure quantum key distribution in quantum cryptography~\cite{2014StefanoRDis,2020LiuDisResource}, the noisy protocols in quantum communication~\cite{2013Datta}, the interferometric schemes in quantum metrology~\cite{2013_Adesso_Metrology, 2014_Adesso_Metrology, 2019_Metrology}, and the micro- and nanoscale heat flow control in quantum thermodynamics~\cite{AHF_2019_NatCommun_Lutz, Pusuluk_2021_PRR}.

Besides its resourcefulness in quantum technologies, the notion of correlation is central to many contemporary fields and has been extended to quantum chemistry since the early 2000s. In particular, the quantum chemistry version of density matrix renormalization group algorithm~\cite{1999_qChemDMRG, 2000_qChemDMRG} has been optimized based on the numerical methods~\cite{2003_qChemDMRG_MutInfo, 2006_qChemDMRG_MutInfo, 2015_qChemDMRG_MutInfo, 2016_qChemDMRG_MutInfo, 2016_qChemDMRG_MutInfo_2} that quantify the pairwise orbital correlations in terms of the quantum mutual information. Moreover, this information-theoretical quantity also has been proposed to investigate the nature of the chemical bonding in different molecular structures~\cite{2011_ChemBond_MutInfo, 2012_ChemBond_MutInfo, 2013_ChemBond_MutInfo, 2013_ChemBond_MutInfo_2, 2014_ChemBond_MutInfo, 2017_ChemBond_MutInfo, 2018_ChemBond_MutInfo}. However, although the quantum mutual information is a measure of the total correlations between two subsystems~\cite{Vedral-2001, Zurek-2002}, the orbital correlations quantified by it have been referred to orbital entanglement until recently.

Ref.~\cite{2021_JCTC_DMRG_SSR} attempted to separate orbital-orbital correlations into classical and quantum parts for the first time in the literature. However, the classical orbital correlations were explored by a distance-based measure, which was first proposed by one of us in Ref.~\cite{Vedral-2001} and quantifies both classical and quantum correlations excluding the entanglement~\cite{2010_PRL_VV_CorrWithRelEnt}. Here, we present a proper decomposition of the total orbital correlations into its classical and quantum parts using the original definition of quantum discord~\cite{Vedral-2001, Zurek-2002} that depends on the difference between two different quantum generalizations of mutual information. Also, we demonstrate whether orbital discord includes orbital entanglement or not by checking the fermionic version of the logarithmic entanglement negativity~\cite{2017_PRB_PTandNegativity, 2017_PRL_PTandNegativity, 2018_PRB_PTandNegativity, 2019_PRA_PTandNegativity}.

The paper is organized as follows. After giving a brief introduction to the fermionic information theory, describing the orbitals as two-mode subsystems in Sec.~\ref{Sec:Methods:PTrace} and discussing the superselection rules imposed on them in Sec.~\ref{Sec:Methods:SSR}, we will present the quantification of orbital discord explicitly in Sec.~\ref{Sec:Methods:Discord}. Section~\ref{Sec:Methods:PTranspose} outlines the quantification of orbital entanglement based on the fermionic partial transpose. The quantum discord and entanglement shared between the orbitals of some prototypical molecules are identified in Sec.~\ref{Sec:Results} We discuss our main results and take a look to future research directions in Sec.~\ref{Sec:Disc}. We conclude with a summary in Sec.~\ref{Sec:Conc}.

\section{Methods} \label{Sec:Methods}

In this part of the paper, we will detail the concepts and methods that are essential for our investigation of the correlations shared between the Hartree-Fock (HF) molecular orbitals (MOs) of the water molecule \(\text{H}_2\text{O}\), 2-propenyl \(\text{C}_3\text{H}_5\), and dicarbon anion \(\text{C}_2^{-}\). The geometries of all these prototypical molecules were optimized at HF/STO-6G level of theory, and they were assumed to be prepared in their electronic ground states. Specifically, we examined the \textit{ab initio} ground states calculated by using the configuration interaction (CI) method~\cite{CI_77, CI_80, CI_81} and considered only the single and double excitations (CISD) above the HF reference state as below:
\begin{equation}\label{Eq:Psi_CISD}
	|\Psi\rangle = \frac{1}{N}\bigg(\mathbb{I} + \sum_{i,a} c_i^a \, f_a^\dagger f_i + \sum_{i>j,a>b} c_{i,j}^{a,b} \, f_a^\dagger f_b^\dagger f_i f_j \bigg) |\Psi_{\mathrm{HF}}\rangle.
\end{equation}
In this above equation, \(N\) is the normalization constant, \(\{c\}\) are the coefficients optimized according to the Rayleigh-Ritz variational principle in the Gaussian09 programme suite~\cite{g09}, where the subscripts and superscripts of these coefficients stand respectively for the occupied and vacant spin-orbitals (fermionic modes) in the reference state, and \(f_\mu^\dagger\) (\(f_\mu\)) is the operator that creates (annihilates) an electron in the $\mu$th mode and obeys the fermionic anticommutation relations
\begin{eqnarray}\label{Eq:Anticomm}
	\{ f_{\mu}, f_{\nu} \} = \{ f_{\mu}^{\dagger}, f_{\nu}^{\dagger} \} = 0, \; \{ f_{\mu}, f_{\nu}^{\dagger} \} = \delta_{\mu \nu} .
\end{eqnarray}

\subsection{Reduced States of Orbitals} \label{Sec:Methods:PTrace}

Each separate term in the expanded form of Eq.~\eqref{Eq:Psi_CISD} is known as an ``configuration'' and corresponds to a Slater determinant, which is an antisymmetrized product of HF spin-orbitals \(\{\psi_\mu\}\). Assuming there are $n_e$ electrons and \(n\) MOs in the molecule, the dominant configuration \(|\Psi_{\mathrm{HF}}\rangle\) can be written in the \(2n\)-mode fermionic Fock space as
\begin{eqnarray}\label{Eq:Psi_HF}
	|\Psi_{\mathrm{HF}}\rangle &=& f_{n_e}^\dagger \cdot\cdot\cdot f_2^\dagger f_1^\dagger \|\Omega\rangle\!\rangle \nonumber \\
	&=& \|\psi_{n_e} \rangle\!\rangle \wedge \cdot\cdot\cdot \wedge \|\psi_2 \rangle\!\rangle \wedge \|\psi_1 \rangle\!\rangle \nonumber \\
	&\equiv& \|1 \cdot\cdot\cdot 1 0 \cdot\cdot\cdot 0\rangle\!\rangle_{\psi_1 \cdot\cdot\cdot \psi_{n_e} \psi_{n_e + 1} \cdot\cdot\cdot \psi_{2 n}} ,
\end{eqnarray}
where \(\psi_{2 \mu - 1}\)and \(\psi_{2 \mu}\) are the \(\mu\)th spin-up and spin-down HF orbitals, the double-lined Dirac notation \(\|\cdot \rangle\!\rangle\) denotes the states in the fermionic Fock space, \(\|\Omega\rangle\!\rangle\) represents the vacuum state, and the wedge product \(\wedge\) provides the antisymmetrization imposed by the anti-commutation relations, e.g.,
\begin{eqnarray}\label{Eq:Wedge}
	\|\psi_{\mu} \rangle\!\rangle \wedge \|\psi_{\nu}\rangle\!\rangle &=& \frac{1}{\sqrt{2}} \big(\|\psi_{\mu}\rangle\!\rangle \otimes \|\psi_{\nu}\rangle\!\rangle - \|\psi_{\nu}\rangle\!\rangle \otimes \|\psi_{\mu}\rangle\!\rangle\big) \nonumber \\
	&=& - \|\psi_{\nu} \rangle\!\rangle \wedge \|\psi_{\mu}\rangle\!\rangle .
\end{eqnarray}
Using the occupation number representation introduced in Eq.~\eqref{Eq:Psi_HF}, we rewrote the post-HF ground state \eqref{Eq:Psi_CISD} as
\begin{eqnarray} \label{Eq:Psi_CISD2}
	|\Psi\rangle &=& \sum_{\vec{s}} \lambda_{\vec{s}} \, \big(f_{2 n}^\dagger\big)^{s_{2 n}} \cdot\cdot\cdot \big(f_{2}^\dagger\big)^{s_{2}} \big(f_{1}^\dagger\big)^{s_{1}}  \|\Omega\rangle\!\rangle \nonumber \\
	&=& \sum_{\vec{s}} \lambda_{\vec{s}} \, \|\ s_1 s_2 \cdot\cdot\cdot s_{2n}\rangle\!\rangle_{\psi_1 \psi_2 \cdot\cdot\cdot \psi_{2 n}} ,
\end{eqnarray}
with \(\vec{s} = \{s_1, s_2, \cdot\cdot\cdot, s_{2n}\}\), \(\sum_{\mu=1}^{2 n} s_\mu = n_e\), and \(s_\mu \in \{0, 1\}\). Here, $\lambda_{\vec{s}} = (-1)^k c(\vec{s}) / N$ where $k$ equals to $n_e - i$ and $i + j$ for the single and double excitation configurations, respectively. Then, we constructed the following $2^{2 n} \times 2^{2 n}$ density matrix
\begin{eqnarray}\label{Eq:Rho_CISD}
	\rho &=& |\Psi\rangle \langle \Psi | \nonumber \\
	&=& \sum_{\vec{s}, \vec{r}} \lambda_{\vec{s}} \lambda_{\vec{r}} \, \big(f_{2 n}^\dagger\big)^{s_{2 n}} \cdot\cdot\cdot \big(f_{1}^\dagger\big)^{s_{1}}  \|\Omega\rangle\!\rangle \langle\!\langle \Omega\| \big(f_{1}\big)^{r_{1}} \cdot\cdot\cdot \big(f_{2 n}\big)^{r_{2 n}} \nonumber \\
	&=& \sum_{\vec{s}, \vec{r}} \lambda_{\vec{s}} \lambda_{\vec{r}} \, \|\ s_1 \cdot\cdot\cdot s_{2n}\rangle\!\rangle \langle\!\langle r_1 \cdot\cdot\cdot r_{2n}\|_{\psi_1 \psi_2 \cdot\cdot\cdot \psi_{2 n}} ,
\end{eqnarray}
and stored only its non-zero elements for further calculations using the triplet sparse matrix representation below
\begin{equation} \label{Eq:Rho_CISD2}
	\rho = \{\lambda_{\vec{s}} \lambda_{\vec{r}}, \vec{s}, \vec{r}\}.
\end{equation}
Next, single and two-orbital reduced states were calculated by taking the partial trace of the state \(\rho\) given in Eq.~\eqref{Eq:Rho_CISD} over the modes \(M = \{\mu_1, \cdot\cdot\cdot,\mu_M\}\) of the remaining orbitals. To this end, we exploited the \textit{inside out} fermionic partial trace operation \cite{2013_PRA_fPTrace, 2021_arXiv_fPTrace} which is given by
\begin{equation} \label{Eq:PTrace1}
	\sigma = \mathrm{tr}_M[\rho] = \mathrm{tr}_{\mu_1} \circ \mathrm{tr}_{\mu_2} \circ \cdot\cdot\cdot \circ \mathrm{tr}_{\mu_M}[\rho] ,
\end{equation}
where the single-mode partial trace operation reads
\begin{eqnarray}\label{Eq:PTrace2}
	& \mathrm{tr}_{\mu}\big[\|\ s_1 \cdot\cdot\cdot s_{2n} \rangle\!\rangle \langle\!\langle r_1 \cdot\cdot\cdot r_{2n}\| \big] & \nonumber \\
	& \equiv\delta_{s_\mu,r_\mu} (-1)^{k^\prime} \|\ s_1 \cdot\cdot\cdot s_{2n} \rangle\!\rangle \langle\!\langle r_1 \cdot\cdot\cdot r_{2n}\|_{\psi_1 \cdot\cdot\cdot \psi_{\mu-1} \psi_{\mu+1} \cdot\cdot\cdot \psi_{2 n}},& \
\end{eqnarray}
with \(k^\prime = s_\mu \sum_{\nu < \mu} s_\nu + r_\mu \sum_{\nu < \mu} r_\nu\).

\subsection{Superselection Rules}\label{Sec:Methods:SSR}

Both the qubits and fermionic modes are two-level distinguishable quantum systems. However, the Jordan-Wigner transformation that maps fermions into qubit systems leads to some ambiguities arising from the absence of a consistent subsystem definition~\cite{FixedN_2003_PRA, PhysRevA.83.062323, 2013_PRA_fPTrace, arXiv_1610.00539}. This is why we need to modify the partial trace operation for fermions by a phase factor in Eq.~(\ref{Eq:PTrace2}). Besides, the superselection rules (SSRs) place additional constraints on the fermionic state space~\cite{PhysRev.88.101}.  Although any superposition in the form of \(\sum_{\vec{s}} \lambda_{\vec{s}} \, | s_1 \cdot\cdot\cdot s_{2n}\rangle\) is a valid quantum state for qubit systems, the nature does not allow a fermionic system to exist in all possible superpositions \(\sum_{\vec{s}} \lambda_{\vec{s}} \, \| s_1 \cdot\cdot\cdot s_{2n}\rangle\!\rangle\). This is related to the conservation of some physical quantities Q by the systems under consideration. Mathematically, Q-SSR forces any valid quantum state \(\rho\) to be block-diagonal with respect to the operator \(\hat{\mathrm{Q}}\), i.e., \([\rho, \hat{\mathrm{Q}}] = 0\).

Parity (P) and particle number (N) are the most relevant observables for the ground state of common molecular systems, i.e., \(\hat{\mathrm{Q}} = \{\hat{\mathrm{P}}, \hat{\mathrm{N}}\}\). As both quantities are globally conserved, Eq.~\eqref{Eq:Rho_CISD} respects total P-SSR and total N-SSR. But they are not conserved locally on each MO in this ground state.

A set of local quantum operations aiming at extracting the nonlocal correlations between MOs cannot change the parity or particle number on single MOs unless the operations are synchronized using classical communication. The orbital correlations become inaccessible even in this case if they stem from the violation of the local conservation laws. Although it may be possible to unlock such correlations transferring them either to a different set of fermionic modes or to an entirely different system by means of global operations, it is reasonable to investigate the effect of local SSRs on orbital correlations. To this goal, we projected the ground state density matrix~(\ref{Eq:Rho_CISD}) into the different sectors of the Fock space as follows
\begin{equation} \label{Eq:Rho_SSR}
	\rho^Q = \sum_{i_1, \cdot\cdot\cdot, i_n} \Pi_{i_n} \wedge \cdot\cdot\cdot \wedge \Pi_{i_1} \rho \, \Pi_{i_1} \wedge \cdot\cdot\cdot \wedge \Pi_{i_n} ,
\end{equation}
where the projectors \(\{\Pi_{i_\mu}\}\) split the $\mu$th single-MO Fock space into its even and odd sectors for local P-SSR, while they separate no excitation, single-excitation and double-excitation sectors inside the same 2-mode Fock spaces for local N-SSR. To illustrate this, let us consider the 2-orbital ground state
\begin{equation}
	\lambda_{1} \|11,00\rangle\!\rangle + \lambda_{2} \|10,01\rangle\!\rangle + \lambda_{3} \|01,10\rangle\!\rangle + \lambda_{4} \|00,11\rangle\!\rangle ,
\end{equation}
where the modes of the two MOs are separated by a comma. Local P-SSR eliminates the density matrix terms \(\rho_{\mu,\nu} = \rho_{\nu,\,\mu} = \lambda_\mu \lambda_\nu\) with \(\mu = \{1, 4\}\) and \(\nu = \{2, 3\}\). The term \(\rho_{1,4} = \rho_{4,1}\) further vanishes after the application of local N-SSR.

\subsection{Classical and Quantum Correlations}\label{Sec:Methods:Discord}

We investigated the pairwise orbital correlations by using two different definitions of the quantum mutual information \cite{Vedral-2001, Zurek-2002}. The amount of the total correlations shared between any two MOs was measured by
\begin{equation} \label{Eq:I}
	I = S(\rho_{L}) + S(\rho_{R}) - S(\rho_{LR}),
\end{equation}
where \(L\) (\(R\)) denoted the left (right) orbital. Here, in Eq.~\eqref{Eq:I}, \(S(\rho)=- \mathrm{tr}[\rho \, \log_2 \rho]\) is the von Neumann entropy. Quantum generalization of an alternative definition of the mutual information was used to quantify the classical correlations
\begin{eqnarray} \label{Eq:C}
	C(R) &=& \max_{\{\Pi_{i_R}\}} \bigg(S(\rho_L) - S(\rho_R|\{\Pi_{i_R}\})\bigg) \nonumber \\
	&=& \max_{\{\Pi_{i_R}\}} \bigg(S(\rho_L) - \sum_{i_R} p_{i_R} S(\rho_{L|\Pi_{i_R}}) \bigg).
\end{eqnarray}
Here, \(S(\rho_L|\{\Pi_{i_R}\})\) is the quantum conditional entropy
of the left orbital, given the complete set of measurements
\(\{\Pi_{i_R}\}\) on the right orbital. The state  $\rho_{L|\Pi_{i_R}}
= \mathrm{tr}_{R} [\Pi_{i_R} \, \rho_{LR} \, \Pi_{i_R}]/p_{i_R}$ is the
post-measurement state of the left orbital with corresponding
probability $p_{i_R} = \mathrm{tr}_{LR} [\Pi_{i_R} \, \rho_{LR}]$.
Physically, Eq.~\eqref{Eq:C} identifies the amount of the maximum information about the left orbital that can be extracted after the measurements \(\{\Pi_{i_R}\}\) performed on the right orbital and called as ``right-classical'' correlation. Similarly, when the shared correlations between the orbitals are extracted by performing local measurements on the left orbital one can end up with ``left-classical'' correlation. It is obvious that \(C(R) \neq C(L)\) in general.

We then quantified the quantum correlations between the orbitals $L$ and $R$, where it is given by the difference between two different quantum generalizations of mutual information given by Eqs.~\eqref{Eq:I} and \eqref{Eq:C}:
\begin{eqnarray} \label{Eq:D}
	D(R) &=& I - C(R) \nonumber \\
	&=& S(\rho_R) - S(\rho_{LR}) + \min_{\{\Pi_{i_R}\}} \bigg(\sum_{i_R} p_{i_R} S(\rho_{L|\Pi_{i_R}})\bigg).
\end{eqnarray}
By definition, quantum discord is asymmetric under the change \(L \leftrightarrow R\). In simple terms, right and left quantum discords --- \(D(R)\) and \(D(L)\), respectively --- do not necessarily reveal the same amount of quantum correlation since the definition of the conditional entropy involves a measurement on one orbital (in Eq.\eqref{Eq:D} right orbital). We calculated both right and left quantum discord.

Let us elaborate on the optimization used in the proper decomposition of the total correlations into classical and quantum parts. The number of parameters required in this optimization is greatly reduced in the presence of SSRs. P-SSR only allows the measurements $\{\Pi_{i_R}\}$ performed in the basis
\begin{eqnarray} \label{Eq:D_PSSR}
	\{\|\varphi\rangle\!\rangle^{\mathrm{P-SSR}}_{\psi_{R_1}\psi_{R_2}}\} \supset
	\begin{cases}
		\alpha_1 \|00\rangle\!\rangle + \alpha_2 \|11\rangle\!\rangle , & \mbox{} \\
		\alpha_3 \|01\rangle\!\rangle + \alpha_4 \|10\rangle\!\rangle , & \mbox{} \\
		\alpha_4^* \|01\rangle\!\rangle - \alpha_3^* \|10\rangle\!\rangle , & \mbox{} \\
		\alpha_2^* \|00\rangle\!\rangle - \alpha_1^* \|11\rangle\!\rangle , & \mbox{}
	\end{cases}
\end{eqnarray}
which means that the optimization involves eight real parameters. On the other hand, N-SSR is conserved only if the measurements $\{\Pi_{i_R}\}$ are carried out in the basis
\begin{eqnarray} \label{Eq:D_NSSR}
	\{\|\varphi\rangle\!\rangle^{\mathrm{N-SSR}}_{\psi_{R_1}\psi_{R_2}}\} \supset
	\begin{cases}
		\qquad \;\;\; \|00\rangle\!\rangle , & \mbox{} \\
		\beta_1 \|01\rangle\!\rangle + \beta_2 \|10\rangle\!\rangle , & \mbox{} \\
		\beta_2^* \|01\rangle\!\rangle - \beta_1^* \|10\rangle\!\rangle , & \mbox{} \\
		\qquad \;\;\; \|11\rangle\!\rangle , & \mbox{}
	\end{cases}
\end{eqnarray}
which in turn decreases the number of optimized real parameters down to four.

\subsection{Quantum Entanglement}\label{Sec:Methods:PTranspose}

Quantum discord includes the quantum entanglement as a subset. Whenever the state is inseparable, the ratio of \(D\) to \(I\) is the only true measure of how much of the total correlation is quantum entanglement. However, quantum correlations are not limited to quantum entanglement and separable mixed states can also possess nonclassical correlations. Discord covers such quantum correlations as well. That is to say, some orbital pairs can still display quantum correlations even they are not entangled.
\begin{table*}[]
\caption{Classical and quantum correlations between Hartree-Fock orbital pairs in the ground state of H$_2$O. Here, $I$ represents the exact value of the total correlation without SSR and $I_{(P/N)}$ equals to its fraction remaining in the presence of P/N-SSR. Both are quantified by the mutual information but only the latter is further decomposed into classical correlation ($C$)~\cite{Vedral-2001} and quantum discord ($D$)~\cite{Vedral-2001, Zurek-2002}. The measurements performed on the left ($L$) and right ($R$) orbitals give the same amount of correlations in the presence of N-SSR. Quantum entanglement $E$ and its fractions $E_{(P/N)}$ in the presence of P/N-SSR are quantified by the fermionic entanglement negativity~\cite{2017_PRB_PTandNegativity, 2017_PRL_PTandNegativity, 2018_PRB_PTandNegativity, 2019_PRA_PTandNegativity}.}
\centering
\begin{ruledtabular}
\begin{tabular}{c | l | r c c | c c r | l r r|}
\multicolumn{11}{c}{H$_2$O \: CISD/STO-6G} \\ [.3ex] \hline
$(L,R)$ & $\qquad I$ & \multicolumn{3}{c|}{$I_{(P)} \;=\quad C_{(P)}(L,R) \quad+\quad D_{(P)}(L,R)$} & \multicolumn{3}{c|}{$I_{(N)} \;=\;\; C_{(N)} \quad+\quad D_{(N)}$} &
$E$ & $E_{(P)}$ & $E_{(N)}$  \\ [.3ex] \hline
$(2,3)$ & $0.21\times 10^{-1}$  & $100{\%}$ & $73.8{\%} , 73.8{\%} $   & $26.2{\%} , 26.2{\%}$   & $84.1{\%}$   & $73.7{\%}$   & $10.4{\%}\quad$   & $0.35\times 10^{-2}$  & $100{\%}$  & $99.3{\%}$   \\
$(2,4)$ & $0.32\times 10^{-1}$  & $56.9{\%}$ & $35.7{\%} ,35.7{\%} $   & $21.2{\%} , 21.2{\%}$   & $42.6{\%}$   & $35.6{\%}$   & $7.0{\%}\quad$   & $0.92\times 10^{-2}$  & $34.8{\%}$  & $34.6{\%}$   \\
$(2,5)$ & $0.19\times 10^{-2}$  & $100{\%}$ & $0.8{\%} , 0.5{\%} $   & \hl{$\textbf{99.2}{\%} , \textbf{99.5}{\%}$}   & $0.37{\%}$   & $0.37{\%}$   & $0{\%}\quad$   & $0.28\times 10^{-5}$  & $100{\%}$  & $0{\%}$   \\
$(2,6)$ & $0.48\times 10^{-1}$  & $99.4{\%}$ & $76.3{\%} , 74.3{\%} $   & $23.1{\%} , 25.1{\%}$    & $75.4{\%}$   & $74.3{\%}$   & $1.0{\%}\quad$ & $0.81\times 10^{-1}$  & $98.2{\%}$  & $0.4{\%}$ \\
$(2,7)$ & $0.19\times 10^{-1}$  & $100{\%}$ & $82.9{\%} , 87.5{\%} $   & $17.1{\%} , 12.5{\%}$   & $80.3{\%}$   & $77.8{\%}$   & \hl{$\underline{\textbf{2.5}}{\%}$}$\quad$   & $0.33\times 10^{-1}$  & $100{\%}$  & \hl{$\underline{\textbf{0}}{\%}$}   \\
$(3,4)$ & $0.94\times 10^{-1}$  & $100{\%}$ & $79.4{\%} , 79.3{\%} $   & $20.6{\%} , 20.7{\%}$   & $88.1{\%}$   & $79.1{\%}$   & $9.0{\%}\quad$   & $0.13\times 10^{-1}$  & $100{\%}$  & $98.9{\%}$   \\
$(3,5)$ & $0.20\times 10^{-2}$  & $100{\%}$ & $3.1{\%} , 1.5{\%} $   & \hl{$\textbf{96.9}{\%} , \textbf{98.5}{\%}$}   & $1.3{\%}$   & $1.3{\%}$   & $0{\%}\quad$   & $0.13\times 10^{-4}$  & $100{\%}$  & $0{\%}$   \\
$(3,6)$ & $0.13$  & $100{\%}$ & $87.8{\%} , 87.6{\%} $   & $12.2{\%} , 12.4{\%}$   & $83.9{\%}$   & $82.4{\%}$   & \hl{$\underline{\textbf{1.5}}{\%}$}$\quad$   & $0.10$  & $100{\%}$  & \hl{$\underline{\textbf{0}}{\%}$}   \\
$(3,7)$ & $0.23$  & $99.9{\%}$ & $72.8{\%} , 73.3{\%} $   & $27.1{\%} , 26.6{\%}$   & $70.7{\%}$   & $69.8{\%}$   & $0.9{\%}\quad$  & $0.24$  & $93.9{\%}$  & $0.4{\%}$   \\
$(4,5)$ & $0.29\times 10^{-2}$  & $100{\%}$ & $1.9{\%} , 1.0{\%} $   & \hl{$\textbf{98.1}{\%} , \textbf{99.0}{\%}$}    & $0.7{\%}$   & $0.7{\%}$   & $0{\%}\quad$   & $0.12\times 10^{-4}$  & $100{\%}$  & $0{\%}$   \\
$(4,6)$ & $0.14$  & $98.1{\%}$ & $78.0{\%} , 74.3{\%} $   & $20.1{\%} , 23.8{\%}$   & $75.5{\%}$   & $74.3{\%}$   &  $1.2{\%}\quad$   & $0.21$  & $69.3{\%}$  &  $0.5{\%}$    \\
$(4,7)$ & $0.10$  & $100{\%}$ & $81.2{\%} , 83.5{\%} $   & $18.8{\%} , 16.5{\%}$  & $77.3{\%}$   & $75.8{\%}$   & \hl{$\underline{\textbf{1.5}}{\%}$}$\quad$   & $0.11$  & $100{\%}$  & \hl{$\underline{\textbf{0}}{\%}$} \\
$(5,6)$ & $0.11\times 10^{-1}$  & $100{\%}$ & $37.7{\%} , 62.0{\%} $   & \hl{$\textbf{62.3}{\%}$}, $38.0{\%}$   & $33.9{\%}$   & $33.9{\%}$   & $0{\%}\quad$   & $0.58\times 10^{-1}$  & $100{\%}$  & $0{\%}$   \\
$(5,7)$ & $0.20\times 10^{-2}$  & $100{\%}$ & $38.3{\%} , 61.2{\%} $   & \hl{$\textbf{61.7}{\%}$}, $38.8{\%}$    & $16.0{\%}$   & $16.0{\%}$   & $0{\%}\quad$   & $0.17\times 10^{-1}$  & $100{\%}$  & $0{\%}$   \\
$(6,7)$ & $0.13$  & $100{\%}$ & $85.1{\%} , 85.1{\%} $   & $14.9{\%} , 14.9{\%}$   & $93.1{\%}$   & $85.0{\%}$   & $8.1{\%}\quad$   & $0.17\times 10^{-1}$  & $100{\%}$  & $99.0{\%}$   \\ [1ex] 
\end{tabular}
\end{ruledtabular}
\label{Table_H2O_pairs}
\end{table*}
\begin{figure*}
 \centering
 \includegraphics[width=\textwidth]{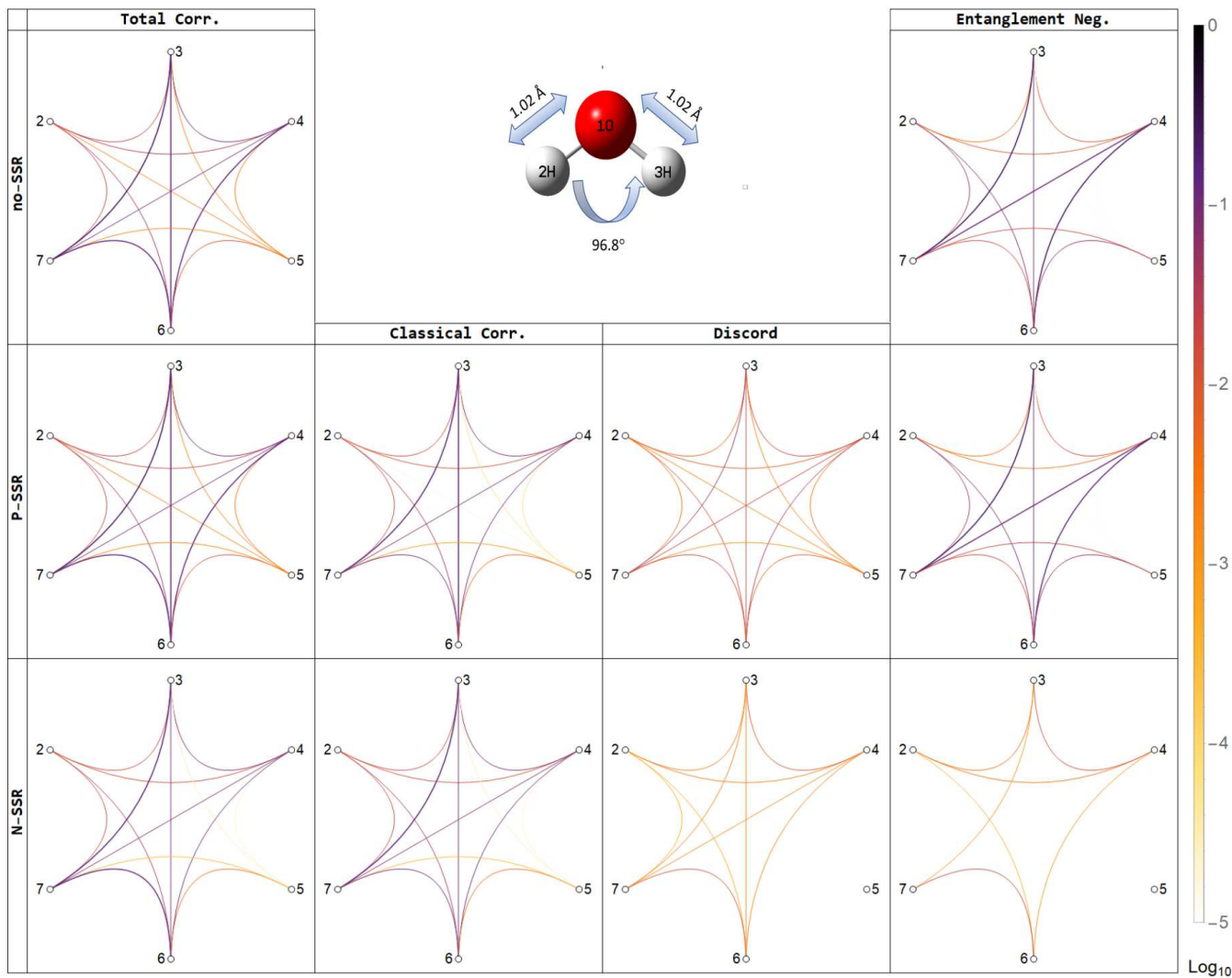}
 \caption{The pairwise total correlations in the singlet ground state of H$_2$O and their proper decomposition into classical correlations and quantum discord in the presence of superselection rules (SSRs). The quantum discord is simply quantum entanglement for the pairs that share nonzero logarithmic negativity on the left panel. Otherwise, it quantifies quantum correlations beyond entanglement.}
\label{Fig::H2O}
\end{figure*}

To check the separability of a discordant orbital pairs, we exploited the fermionic partial transpose proposed and investigated in Refs.~\cite{2017_PRB_PTandNegativity, 2017_PRL_PTandNegativity, 2018_PRB_PTandNegativity, 2019_PRA_PTandNegativity}. Likewise the fermionic partial trace in Eq.~\eqref{Eq:PTrace2}, the most foundational distinction that separates the fermionic partial transpose from the qubit partial transpose is a phase factor as below:
\begin{eqnarray} \begin{aligned} \label{Eq:PTranspose}
		\big(\|\ s_{L_1} \cdot\cdot\cdot s_{L_{\mu}}, s_{R_1} \cdot\cdot\cdot s_{R_\nu} &\rangle\!\rangle \langle\!\langle r_{L_1} \cdot\cdot\cdot r_{L_\mu}, r_{R_1} \cdot\cdot\cdot r_{R_\nu}\| \big)^{\mathrm{T}_R} \\
		&\,\shortmid\shortmid\shortmid \\
		(-1)^\phi \hat{\mathrm{U}}_R \|\ s_{L_1} \cdot\cdot\cdot s_{L_{\mu}}, r_{R_1} \cdot\cdot\cdot r_{R_\nu} &\rangle\!\rangle \langle\!\langle r_{L_1} \cdot\cdot\cdot r_{L_\mu}, s_{R_1} \cdot\cdot\cdot s_{R_\nu}\| \hat{\mathrm{U}}^\dagger_R ,
\end{aligned} \end{eqnarray}
where
\begin{equation}
	\phi = [(\tau_R + \bar{\tau}_R) \mathrm{mod 2}]/2 + (\tau_R + \bar{\tau}_R)(\tau_L + \bar{\tau}_L) ,
\end{equation}
with \(\tau_{L(R)} = \sum_{j=L_1(R_1)}^{L_\mu(R_\nu)} s_{j}\) , \(\bar{\tau}_{L(R)} = \sum_{j=L_1(R_1)}^{L_\mu(R_\nu)} r_{j}\), and \(\hat{\mathrm{U}}_R = \prod_{j=R_1}^{R_\nu} (f_j + f^\dagger_j)\). As we focused on the pairwise orbital correlations, we considered only the bipartitions $L|R$ defined by $L_\mu = R_\nu = 2$ in this paper. Then we calculated the fermionic logarithmic negativity proposed and investigated in Refs.~\cite{2017_PRB_PTandNegativity, 2017_PRL_PTandNegativity, 2018_PRB_PTandNegativity, 2019_PRA_PTandNegativity}, that reads
\begin{equation} \label{Eq:E}
	E = \log_2 \|\rho_{LR}^{T_R}\| ,
\end{equation}
where $\|\rho\|$ denotes the trace norm that equals to \(\mathrm{tr}[\sqrt{\rho \rho^\dagger}]\).

It is important to note that we utilized this measure not to quantify orbital-orbital entanglement but to reveal the nature of the orbital correlations quantified by quantum discord. The amount of quantum discord calculated by Eq.~\eqref{Eq:D} was recognized as quantum entanglement for non-zero values of \(E\). On the other hand, the value determined by \(D\) was regarded as quantum correlations beyond entanglement when \(E\) vanishes.

At this point, we should also emphasize that SSRs are believed to force the fermionic systems not to exist in bound entangled states that cannot be captured by the measure \(E\)~\cite{2019_PRA_PTandNegativity}.

\section{Results} \label{Sec:Results}

In what follows we present the total, classical, and quantum correlations between the pairs of the MOs of the ground states of the water molecule \(\text{H}_2\text{O}\), 2-propenyl \(\text{C}_3\text{H}_5\), and dicarbon anion \(\text{C}_2^{-}\). Although the CISD ground states include all the HF MOs approximated in the STO-6G minimal basis set, we exclude the frozen orbitals while visualizing our data in the next three sections.

\subsection{Water}

As shown in Fig.~\ref{Fig::H2O}, the electronic structure of water molecule \(\text{H}_2\text{O}\) consists of 10-electrons and 7 HF MOs. Here, the \(1 a_1\), \(2 a_1\), \(1 b_2\), \(3 a_1\), \(1 b_1\), \(4 a_1\), and \(2 b_2\) orbitals are indexed in ascending order from 1 to 7. The geometry is defined by the bond length and the bond angle that are \(1.02 \mbox{\normalfont\AA}\) and \(96.8\degree\), respectively, and the CISD energy is converged to \(-75.737545702 \, E_\mathrm{h}\).

Details of the numerical results are given in Table \ref{Table_H2O_pairs}. There, the exact values of the total correlation \(I\) and entanglement \(E\) without SSR accompany their fractions remaining in the presence of SSRs. Further, Table \ref{Table_H2O_pairs} contains what percentage of the total correlation decomposes into classical correlation and quantum discord in the presence of the local P-SSR and N-SSR.

We have two particularly noteworthy results that should be mentioned here. First, even though the classical part tends to predominate the total correlation, there exist some cases with quantum discord as the dominant correlation in the presence of the local P-SSR. The cases in question are the pairs including the \(5\)th orbital \(1 b_1\).

The \(1 b_1\) is the HOMO of the HF ground state and completely formed from the \(2p_x\) electron lone pair of the oxygen atom. However, when the second orbital in the pair is one of the highest MOs indexed by \(6\) and \(7\), the dominance of quantum correlations holds only for the left-discord. That is to say, the quantum correlations can be stronger than the classical correlations for the pairs \((5,6)\) and \((5,7)\) only if the extraction of the latter is performed by the local measurements on the lone-pair orbital \(1 b_1\). Note that \(6\)th and \(7\)th orbitals participating in these pairs are \(4 a_1\) and \(2 b_2\), and have an antibonding character.

Second, after applying the local N-SSR, quantum discord between the pairs \((2,7)\), \((3,6)\), and \((4,7)\) survives even in the absence of entanglement. The \(2\)nd and \(3\)rd orbitals are bonding MOs \(2 a_1\) and \(1 b_2\), while the \(4\)th orbital \(3 a_1\) has a character closer to a lone-pair MO than a bonding MO. Hence, these discordant pairs are composed of occupied and vacant HF MOs that have opposite \(a_1/b_2\) symmetries.

\begin{figure*}
 \centering
 \includegraphics[width=\textwidth]{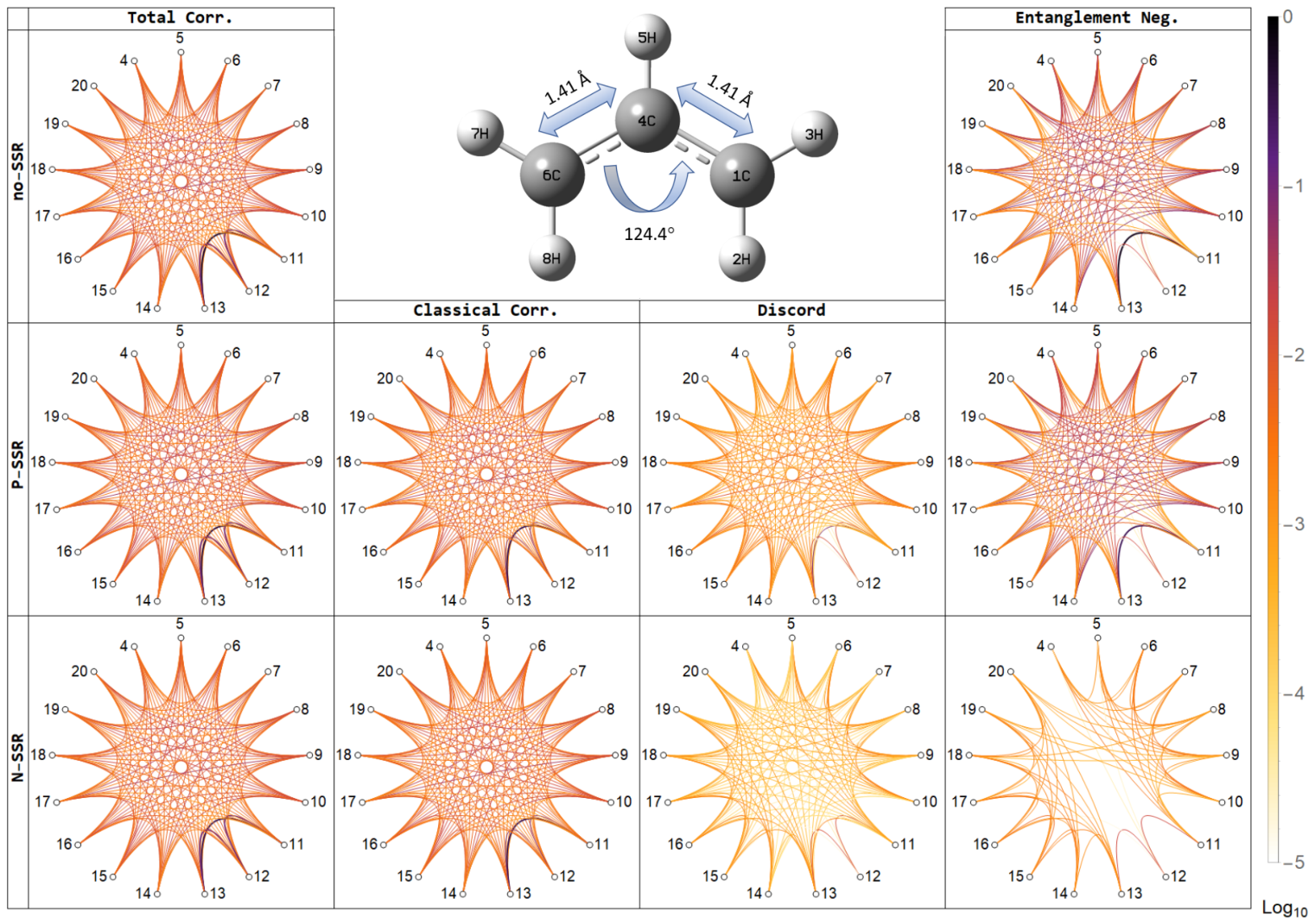}
 \caption{The pairwise total correlations in the doublet ground state of C$_3$H$_5$ and their proper decomposition into classical correlations and quantum discord in the presence of superselection rules (SSRs). The quantum discord is simply quantum entanglement for the pairs that share nonzero logarithmic negativity on the left panel. Otherwise, it quantifies quantum correlations beyond entanglement.}
\label{Fig::C3H5}
\end{figure*}

\begin{figure*}
 \centering
 \includegraphics[width=\textwidth]{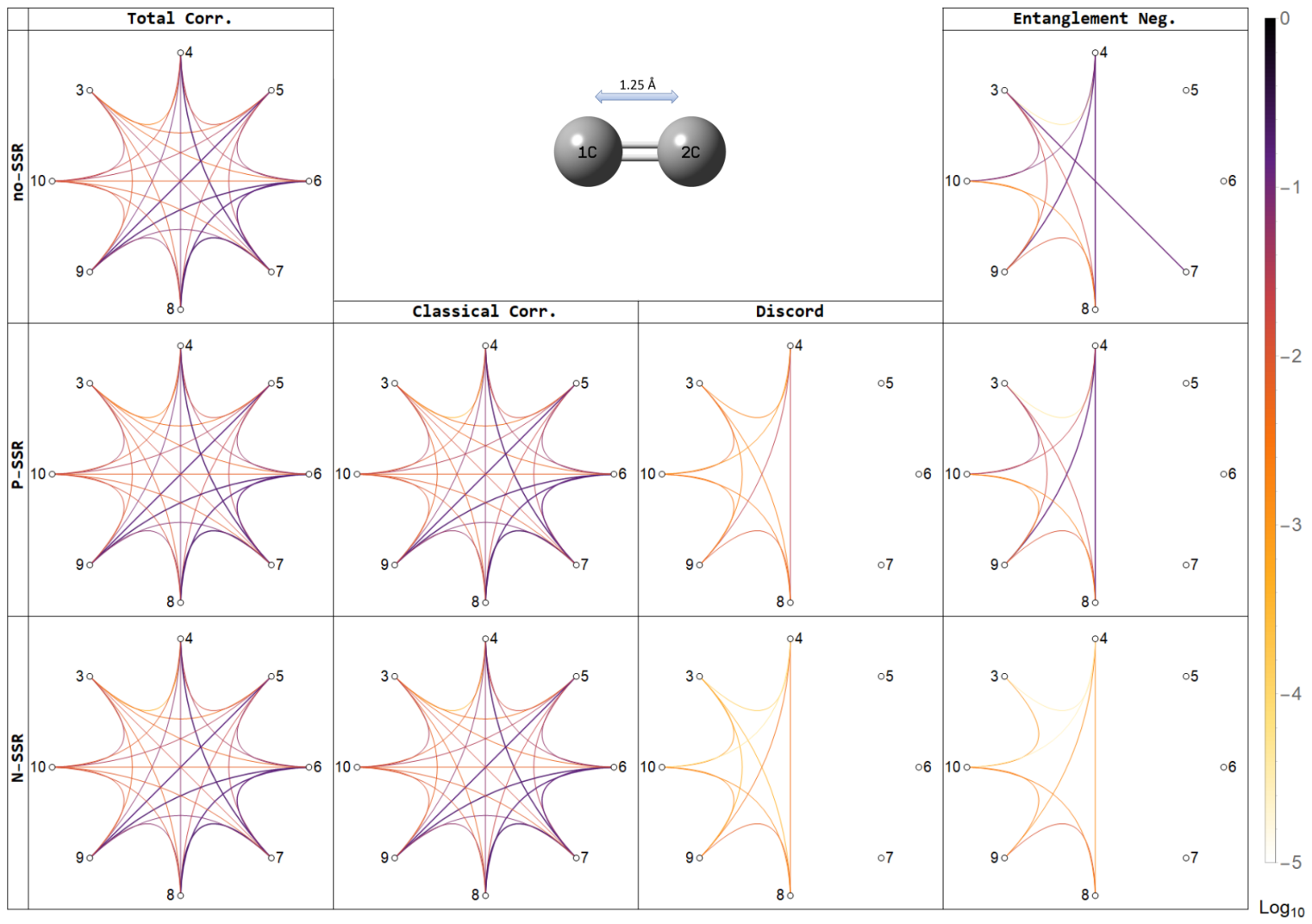}
 \caption{The pairwise total correlations in the doublet ground state of C$_2^-$ and their proper decomposition into classical correlations and quantum discord in the presence of superselection rules (SSRs). The quantum discord is simply quantum entanglement for the pairs that share nonzero logarithmic negativity on the left panel. Otherwise, it quantifies quantum correlations beyond entanglement.}
\label{Fig::C2}
\end{figure*}

\subsection{2-Propenyl}

The bond lengths and angles between the neighboring carbon atoms in \(\text{C}_3\text{H}_5\) are respectively \(1.41 \mbox{\normalfont\AA}\) and \(124.4\degree\) in the geometry optimized with the STO-6G minimal basis set. The electronic structure of the molecule involves \(23\)-electrons and \(20\) HF MOs, with the final energy of \(-116.35450180\, E_\mathrm{h}\).

In Fig.~\ref{Fig::C3H5}, the total correlation without SSRs reaches its maximum value \(0.49\) and minimum value \(0.00097\) for the pairs of orbitals \((11, 13)\) and \((11, 17)\), respectively. Obviously, the classical part of the total correlation is more dominant for all pairs of orbitals.

The quantum entanglement and discord possessed by the pairs of orbitals \((11, 12)\) and \((12, 13)\) are robust against local SSRs. As a matter fact, the orbital \(12\) does not share a non-negligible amount of logarithmic entanglement negativity with those other than \(11\) and \(13\). Also, the pair \((11, 13)\) has quantum discord in the presence of the local N-SSR, though it exist in a separable state. These three orbitals share a unique symmetry different from the rest. Each of them is a superposition of only the three \(2p_z\) orbitals of carbon atoms, and therefore, they display a \(\pi\)-orbital character.
Moreover, they respectively correspond to HOMO, half-filled MO, and LUMO in the HF ground state.

Besides, although they do not have any entanglement with or without SSRs, the pairs of orbitals \((7, 13)\), \((8, 13)\), \((9, 13)\),\((10, 13)\),\((11, 14)\), and \((11, 17)\) always have a nonzero quantum discord. Finally, under the local N-SSR, the ground state of \(\text{C}_3\text{H}_5\) reveals \(51\) other orbital pairs exhibiting quantum discord in the absence of entanglement (more details can be seen in Fig.~\ref{Fig::C3H5}).

\subsection{Dicarbon anion}

The interatomic distance is \(1.25 \mbox{\normalfont\AA}\) in the optimized molecular geometry of the dicarbon anion \(\text{C}_2^{-}\) in its doublet ground state. The electronic structure is composed of \(13\)-electrons and \(10\) orbitals, with the final energy of \(-75.293354472\, E_\mathrm{h}\).

The numerical values of the total correlation without SSRs lie in the range between \(0.00025\) and \(0.2\) corresponding to the pairs of orbitals \((3, 4)\) and \((7, 8)\), respectively. The former is
the only pair whose dominant pairwise correlation is quantum discord in the presence of the local P-SSR. Also, its discord does not vanish under the local N-SSR while its entanglement negativity is always small enough to be neglected. Actually, the \(3\)rd and \(4\)th orbitals that constitute this pair are the HF MOs \(2 \sigma_g\) and \(2 \sigma_u^*\). The simultaneous occupancy of these two orbitals, which has a probability of \(0.91\) in the ground state, implies the absence of a \(\sigma\) bond between the C atoms and reduces the bond order of the molecule.

As shown in Fig.~\ref{Fig::C2}, the orbitals \(5\), \(6\), and \(7\) do not share any quantum correlation with the other ones under the local SSRs. These orbitals are HOMO, half-filled MO, and LUMO in the HF ground state. Besides, each of them is composed of \(\alpha\) and \(\beta\) modes having incompatible symmetries (please see the red, purple, and green arrow pairs in Fig.~\ref{Fig::C2v2}). The \(5\)th and \(7\)th orbitals are mixtures of \(\sigma_g\) and \(\pi_u\) symmetries. The former (latter) includes a \(3 \sigma_g\) \(\alpha\) (\(\beta\)) mode and a \(1 \pi_u\) \(\beta\) (\(\alpha\)) mode. Both modes in the \(6\)th orbital posses \(\pi_u\) symmetry, but the \(\alpha\) and \(\beta\) modes are symmetric superpositions of the two \(2p_y\) and the two \(2p_x\) orbitals of carbon atoms, respectively. Please note that these six \(\alpha\) and \(\beta\) modes are ordered differently in the neutral dicarbon molecule and constitute two degenerate \(1 \pi_u\) orbitals and one \(3 \sigma_g\) orbital, which are HOMOs and LUMO of the molecule, respectively.

Quantum discord in the pairs of orbitals  \((3, 8)\) and \((3, 9)\) exist despite the lack of entanglement in the presence of the local N-SSR. The \(3\)rd orbital is the bonding \(\sigma\) orbital (\(2 \sigma_g\)), while the \(8\)th and \(9\)th ones are degenerate antibonding \(\pi\)-orbitals (\(1 \pi_g^*\)). Furthermore, the pairwise quantum correlations distributed among the antibonding orbitals 8, 9, and 10 seem not to be affected significantly by the local SSRs. These three orbitals
are the unoccupied MOs in the HF ground state. The last one is \(3 \sigma_u^*\).


\section{Discussions} \label{Sec:Disc}

A natural question to ask at this point is that what would happen if the density matrix of an orbital pair described a qubit system? To address this question, we recalculated the entanglement in two-orbital density matrices using the multipartite entanglement measure proposed in Ref.~\cite{2011Taming}. This measure becomes the negativity for two-qubit states and is easily computable using the code provided in Ref.~\cite{pptmixer}, together with the parser YALMIP~\cite{Lofberg2004} and the solver SDPT3~\cite{Tutuncu99, Tutuncu03}. Then it turns out that some density matrices with fermionic mode entanglement are separable for qubit systems (see Supplementary Material for more details). It means that the orbital entanglement would be underestimated if the orbital density matrices were treated as qubit states.
\begin{figure*}
 \centering
 \includegraphics[width=\textwidth]{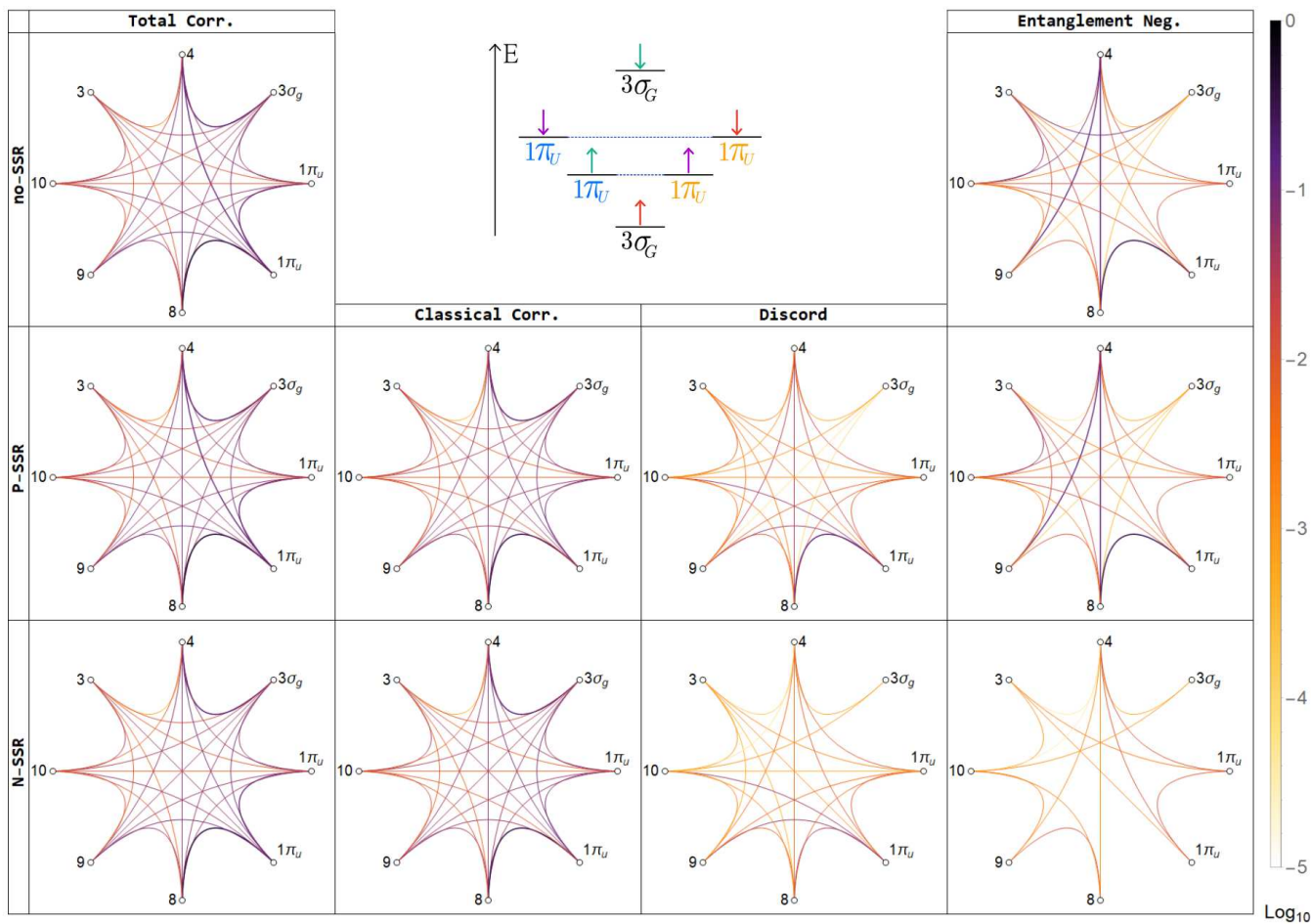}
 \caption{The orbital correlations in the doublet ground state of C$_2^-$ in the case when the mode pairs with the same symmetry are considered as subsystem as in the singlet state of C$_2$. The orbitals \(5\), \(6\), and \(7\) in Fig.~\ref{Fig::C2} consist of the \(\alpha\) and \(\beta\) modes represented by red, purple, and green up and down arrow pairs, respectively. However, the HF MOs in the ground state of C$_2$ are composed of the mode pairs sharing the same symmetry, i.e., they are two degenerate \(1 \pi_u\) orbitals and one \(3 \sigma_g\) orbital highlighted in blue, orange, and black in the energy diagram.}
\label{Fig::C2v2}
\end{figure*}

Another fundamental question that we can ask here is how the chosen orbital basis affects the classical and quantum orbital correlations. Here, we have so far investigated the correlations between HF MOs, which are also known as delocalized or canonical MOs. These orthogonal orbitals can be transformed to the so-called localized MOs by a unitary transformation which makes \(|\Psi_{\mathrm{HF}}\rangle\) defined in Eq.~(\ref{Eq:Psi_HF}) invariant. However, the coefficients of the other terms in CISD ground state~(\ref{Eq:Psi_CISD2}) do not remain the same after this transformation. Hence, the orbital correlations strongly depend on the MOs under consideration. This seems reasonable as a unitary transformation preserves the global properties such as total energy, but not the subsystem structure and the resulting local properties.

For the open shell molecules, there is an ambiguity in the definition of MOs as two-mode subsystems even in the absence of any transformation. This arises from the fact that the unrestricted HF method allows the \(\alpha\) and \(\beta\) modes to have different energies even if they share the same symmetry, i.e., if they can be written as almost the same superposition state in the basis of atomic orbitals. In some molecules such as C\(_2^-\), the situation becomes more complicated. The modes sharing the same symmetry may appear in different orders when the \(\alpha\) and \(\beta\) modes are ordered separately according to their energies. But how would orbital correlations alter if two-mode subsystems were defined by symmetry similarity rather than the energy ordering?

Let us reconsider the orbitals \(5\), \(6\), and \(7\) that do not share any quantum correlation with the other MOs in the dicarbon anion. Assume we redefine these three two-mode subsystems by combining the constituent \(\alpha\) and \(\beta\) modes with the same symmetry. In this case, the resulting orbitals become strongly correlated with the rest as shown in Fig.~\ref{Fig::C2v2}. As a matter of fact, they are equal to the \(5\)th \(1 \pi_u\) orbital, \(6\)th \(1 \pi_u\) orbital, and \(7\)th \(3 \sigma_g\) orbital in the neutral dicarbon molecule. Then it is not clear which orbital definition would be more appropriate to understand what happens to correlations in the transition from the neutral to the anionic state of the molecule. In this respect, we are planning to extend the present methodology to investigate the dynamical nature of orbital correlations during chemical processes.


Although the electronic structure of molecules and many of their observed properties are fairly well described by HF MOs, it should be emphasized that they represent an approximation to reality. If we could solve the electronic Schr\"{o}dinger equations exactly without any approximation, we would get ``real'' molecular ground states. However, these states would still be superpositions in the bases that describe the individual atoms. That is, it would still be possible, at least in principle, to find classical and quantum correlations in the ground states of molecules.


Finally, we would also stress that the local SSRs increase the entropy of molecular electronic states while removing some part of their classical and quantum correlations. For example, the entropy of the water molecule is risen in its ground state approximately by \(0.13\) and \(0.31\) after P-SSR and N-SSR, respectively. The same values are \(0.76\) and \(0.84\) for the dicarbon anion. Thus, although some orbital correlations become inaccessible in the presence of SSRs, the entropic cost of this inaccessibility can reduce the free energy of the molecules. Even the orbital correlations frozen by SSRs can become important for the quantum information processing tasks indirectly in this way. Moreover, these frozen correlations can be also unlocked by transferring them either to a different set of fermionic modes or to an entirely different system by means of global operations. However, the possible resource values of the orbital correlations frozen by local SSRs are beyond the scope of this methodological paper, though their investigation is a natural direction to pursue future work.


\section{Conclusion} \label{Sec:Conc}

In the present work, we have analyzed the electronic ground states of three prototypical molecules, namely \(\text{H}_2\text{O}\), \(\text{C}_3\text{H}_5\), and \(\text{C}_2^{-}\), using the tools of fermionic information theory that is still an active research area.

Although the \(\alpha\) and \(\beta\) modes are two-level distinguishable quantum systems, the map of molecular electronic states into multipartite qubit states leads to some ambiguities in defining the subsystems. Therefore, we have applied the fermionic partial trace operation to obtain the single- and double-orbital states. We have further taken into account the additional constraints that the local (parity and number) superselection rules put on the spaces in which these reduced states live. In particular, we have found that the computational complexity of orbital discord is significantly lower when compared to the four-qubit quantum discord. This enables us to explicitly discuss both the classical and quantum parts of the pairwise orbital correlations in the presence of superselection rules.

Remarkably, we have observed that quantum orbital correlations can dominate the total orbital correlations and survive even in the absence of orbital entanglement. We have also examined whether the symmetries of the constituent orbitals make a difference in terms of correlations. Besides, when the orbital density matrices are treated as qubit states, we have shown that orbital entanglement would be underestimated.

We believe our work can serve the needs of future investigations in understanding the nature of the correlations in chemical systems.


\begin{acknowledgments}

This research was supported by the Scientific and Technological Research Council of Turkey (T\"{U}B\.{I}TAK) under Grant No. (120F089).

\end{acknowledgments}


\begin{thebibliography}{59}%
\makeatletter
\providecommand \@ifxundefined [1]{%
 \@ifx{#1\undefined}
}%
\providecommand \@ifnum [1]{%
 \ifnum #1\expandafter \@firstoftwo
 \else \expandafter \@secondoftwo
 \fi
}%
\providecommand \@ifx [1]{%
 \ifx #1\expandafter \@firstoftwo
 \else \expandafter \@secondoftwo
 \fi
}%
\providecommand \natexlab [1]{#1}%
\providecommand \enquote  [1]{``#1''}%
\providecommand \bibnamefont  [1]{#1}%
\providecommand \bibfnamefont [1]{#1}%
\providecommand \citenamefont [1]{#1}%
\providecommand \href@noop [0]{\@secondoftwo}%
\providecommand \href [0]{\begingroup \@sanitize@url \@href}%
\providecommand \@href[1]{\@@startlink{#1}\@@href}%
\providecommand \@@href[1]{\endgroup#1\@@endlink}%
\providecommand \@sanitize@url [0]{\catcode `\\12\catcode `\$12\catcode
  `\&12\catcode `\#12\catcode `\^12\catcode `\_12\catcode `\%12\relax}%
\providecommand \@@startlink[1]{}%
\providecommand \@@endlink[0]{}%
\providecommand \url  [0]{\begingroup\@sanitize@url \@url }%
\providecommand \@url [1]{\endgroup\@href {#1}{\urlprefix }}%
\providecommand \urlprefix  [0]{URL }%
\providecommand \Eprint [0]{\href }%
\providecommand \doibase [0]{https://doi.org/}%
\providecommand \selectlanguage [0]{\@gobble}%
\providecommand \bibinfo  [0]{\@secondoftwo}%
\providecommand \bibfield  [0]{\@secondoftwo}%
\providecommand \translation [1]{[#1]}%
\providecommand \BibitemOpen [0]{}%
\providecommand \bibitemStop [0]{}%
\providecommand \bibitemNoStop [0]{.\EOS\space}%
\providecommand \EOS [0]{\spacefactor3000\relax}%
\providecommand \BibitemShut  [1]{\csname bibitem#1\endcsname}%
\let\auto@bib@innerbib\@empty
\bibitem [{\citenamefont {{Weinhold}}(1999)}]{1999WeinholdCbonding}%
  \BibitemOpen
  \bibfield  {author} {\bibinfo {author} {\bibfnamefont {F.~A.}\ \bibnamefont
  {{Weinhold}}},\ }\bibfield  {title} {\bibinfo {title} {{Chemical Bonding as a
  Superposition Phenomenon}},\ }\href {https://doi.org/10.1021/ed076p1141}
  {\bibfield  {journal} {\bibinfo  {journal} {J. Chem. Edu.}\ }\textbf
  {\bibinfo {volume} {76}},\ \bibinfo {pages} {1141} (\bibinfo {year}
  {1999})}\BibitemShut {NoStop}%
\bibitem [{\citenamefont {Pauling}(1960)}]{1960_Pauling}%
  \BibitemOpen
  \bibfield  {author} {\bibinfo {author} {\bibfnamefont {L.}~\bibnamefont
  {Pauling}},\ }\href@noop {} {\emph {\bibinfo {title} {The nature of the
  chemical bond and the structure of molecules and crystals.}}},\ \bibinfo
  {edition} {3rd}\ ed.\ (\bibinfo  {publisher} {Cornell University Press},\
  \bibinfo {year} {1960})\BibitemShut {NoStop}%
\bibitem [{\citenamefont {Coulson}(1952)}]{1952_Coulson}%
  \BibitemOpen
  \bibfield  {author} {\bibinfo {author} {\bibfnamefont {C.~A.}\ \bibnamefont
  {Coulson}},\ }\href@noop {} {\emph {\bibinfo {title} {Valance}}},\ \bibinfo
  {edition} {1st}\ ed.\ (\bibinfo  {publisher} {Oxford University Press},\
  \bibinfo {year} {1952})\BibitemShut {NoStop}%
\bibitem [{\citenamefont {Hiberty}\ and\ \citenamefont
  {Leforestier}(1978)}]{1978_MO2VB}%
  \BibitemOpen
  \bibfield  {author} {\bibinfo {author} {\bibfnamefont {P.~C.}\ \bibnamefont
  {Hiberty}}\ and\ \bibinfo {author} {\bibfnamefont {C.}~\bibnamefont
  {Leforestier}},\ }\bibfield  {title} {\bibinfo {title} {Expansion of
  molecular orbital wave functions into valence bond wave functions. a
  simplified procedure},\ }\href {https://doi.org/10.1021/ja00475a007}
  {\bibfield  {journal} {\bibinfo  {journal} {J. Am. Chem. Soc.}\ }\textbf
  {\bibinfo {volume} {49}},\ \bibinfo {pages} {473001} (\bibinfo {year}
  {1978})}\BibitemShut {NoStop}%
\bibitem [{\citenamefont {Shaik}\ \emph {et~al.}(2021)\citenamefont {Shaik},
  \citenamefont {Danovich},\ and\ \citenamefont {Hiberty}}]{2021_MO2VB}%
  \BibitemOpen
  \bibfield  {author} {\bibinfo {author} {\bibfnamefont {S.}~\bibnamefont
  {Shaik}}, \bibinfo {author} {\bibfnamefont {D.}~\bibnamefont {Danovich}},\
  and\ \bibinfo {author} {\bibfnamefont {P.~C.}\ \bibnamefont {Hiberty}},\
  }\bibfield  {title} {\bibinfo {title} {Valence bond theory--its birth,
  struggles with molecular orbital theory, its present state and future
  prospects},\ }\href {https://doi.org/10.3390/molecules26061624} {\bibfield
  {journal} {\bibinfo  {journal} {Molecules}\ }\textbf {\bibinfo {volume}
  {26}},\ \bibinfo {pages} {1624} (\bibinfo {year} {2021})}\BibitemShut
  {NoStop}%
\bibitem [{\citenamefont {MacFarlane}\ \emph {et~al.}(2003)\citenamefont
  {MacFarlane}, \citenamefont {Dowling},\ and\ \citenamefont
  {Milburn}}]{2003_QTech}%
  \BibitemOpen
  \bibfield  {author} {\bibinfo {author} {\bibfnamefont {A.~G.~J.}\
  \bibnamefont {MacFarlane}}, \bibinfo {author} {\bibfnamefont {J.~P.}\
  \bibnamefont {Dowling}},\ and\ \bibinfo {author} {\bibfnamefont {G.~J.}\
  \bibnamefont {Milburn}},\ }\bibfield  {title} {\bibinfo {title} {Quantum
  technology: the second quantum revolution},\ }\href
  {https://doi.org/10.1098/rsta.2003.1227} {\bibfield  {journal} {\bibinfo
  {journal} {Phil. Trans. R. Soc. Lond. A}\ }\textbf {\bibinfo {volume}
  {361}},\ \bibinfo {pages} {1655} (\bibinfo {year} {2003})}\BibitemShut
  {NoStop}%
\bibitem [{\citenamefont {Modi}\ \emph {et~al.}(2012)\citenamefont {Modi},
  \citenamefont {Brodutch}, \citenamefont {Cable}, \citenamefont {Paterek},\
  and\ \citenamefont {Vedral}}]{2012ModiCorrelations}%
  \BibitemOpen
  \bibfield  {author} {\bibinfo {author} {\bibfnamefont {K.}~\bibnamefont
  {Modi}}, \bibinfo {author} {\bibfnamefont {A.}~\bibnamefont {Brodutch}},
  \bibinfo {author} {\bibfnamefont {H.}~\bibnamefont {Cable}}, \bibinfo
  {author} {\bibfnamefont {T.}~\bibnamefont {Paterek}},\ and\ \bibinfo {author}
  {\bibfnamefont {V.}~\bibnamefont {Vedral}},\ }\bibfield  {title} {\bibinfo
  {title} {The classical-quantum boundary for correlations: Discord and related
  measures},\ }\href {https://doi.org/10.1103/RevModPhys.84.1655} {\bibfield
  {journal} {\bibinfo  {journal} {Rev. Mod. Phys.}\ }\textbf {\bibinfo {volume}
  {84}},\ \bibinfo {pages} {1655} (\bibinfo {year} {2012})}\BibitemShut
  {NoStop}%
\bibitem [{\citenamefont {Adesso}\ \emph {et~al.}(2016)\citenamefont {Adesso},
  \citenamefont {Bromley},\ and\ \citenamefont {Cianciaruso}}]{2016AdessoQC}%
  \BibitemOpen
  \bibfield  {author} {\bibinfo {author} {\bibfnamefont {G.}~\bibnamefont
  {Adesso}}, \bibinfo {author} {\bibfnamefont {T.~R.}\ \bibnamefont
  {Bromley}},\ and\ \bibinfo {author} {\bibfnamefont {M.}~\bibnamefont
  {Cianciaruso}},\ }\bibfield  {title} {\bibinfo {title} {Measures and
  applications of quantum correlations},\ }\href
  {https://doi.org/10.1088/1751-8113/49/47/473001} {\bibfield  {journal}
  {\bibinfo  {journal} {J. Phys. A: Math. Theor.}\ }\textbf {\bibinfo {volume}
  {49}},\ \bibinfo {pages} {473001} (\bibinfo {year} {2016})}\BibitemShut
  {NoStop}%
\bibitem [{\citenamefont {Horodecki}\ \emph {et~al.}(2009)\citenamefont
  {Horodecki}, \citenamefont {Horodecki}, \citenamefont {Horodecki},\ and\
  \citenamefont {Horodecki}}]{2009HorodeckiEnt}%
  \BibitemOpen
  \bibfield  {author} {\bibinfo {author} {\bibfnamefont {R.}~\bibnamefont
  {Horodecki}}, \bibinfo {author} {\bibfnamefont {P.}~\bibnamefont
  {Horodecki}}, \bibinfo {author} {\bibfnamefont {M.}~\bibnamefont
  {Horodecki}},\ and\ \bibinfo {author} {\bibfnamefont {K.}~\bibnamefont
  {Horodecki}},\ }\bibfield  {title} {\bibinfo {title} {Quantum entanglement},\
  }\href {https://doi.org/10.1103/RevModPhys.81.865} {\bibfield  {journal}
  {\bibinfo  {journal} {Rev. Mod. Phys.}\ }\textbf {\bibinfo {volume} {81}},\
  \bibinfo {pages} {865} (\bibinfo {year} {2009})}\BibitemShut {NoStop}%
\bibitem [{\citenamefont {Reid}\ \emph {et~al.}(2009)\citenamefont {Reid},
  \citenamefont {Drummond}, \citenamefont {Bowen}, \citenamefont {Cavalcanti},
  \citenamefont {Lam}, \citenamefont {Bachor}, \citenamefont {Andersen},\ and\
  \citenamefont {Leuchs}}]{2009_EntTech}%
  \BibitemOpen
  \bibfield  {author} {\bibinfo {author} {\bibfnamefont {M.~D.}\ \bibnamefont
  {Reid}}, \bibinfo {author} {\bibfnamefont {P.~D.}\ \bibnamefont {Drummond}},
  \bibinfo {author} {\bibfnamefont {W.~P.}\ \bibnamefont {Bowen}}, \bibinfo
  {author} {\bibfnamefont {E.~G.}\ \bibnamefont {Cavalcanti}}, \bibinfo
  {author} {\bibfnamefont {P.~K.}\ \bibnamefont {Lam}}, \bibinfo {author}
  {\bibfnamefont {H.~A.}\ \bibnamefont {Bachor}}, \bibinfo {author}
  {\bibfnamefont {U.~L.}\ \bibnamefont {Andersen}},\ and\ \bibinfo {author}
  {\bibfnamefont {G.}~\bibnamefont {Leuchs}},\ }\bibfield  {title} {\bibinfo
  {title} {Colloquium: The
  $\mathrm{E}$instein-$\mathrm{P}$odolsky-$\mathrm{R}$osen paradox: From
  concepts to applications},\ }\href
  {https://doi.org/10.1103/RevModPhys.81.1727} {\bibfield  {journal} {\bibinfo
  {journal} {Rev. Mod. Phys.}\ }\textbf {\bibinfo {volume} {81}},\ \bibinfo
  {pages} {1727} (\bibinfo {year} {2009})}\BibitemShut {NoStop}%
\bibitem [{\citenamefont {Pezz\`e}\ \emph {et~al.}(2018)\citenamefont
  {Pezz\`e}, \citenamefont {Smerzi}, \citenamefont {Oberthaler}, \citenamefont
  {Schmied},\ and\ \citenamefont {Treutlein}}]{2018_EntTech}%
  \BibitemOpen
  \bibfield  {author} {\bibinfo {author} {\bibfnamefont {L.}~\bibnamefont
  {Pezz\`e}}, \bibinfo {author} {\bibfnamefont {A.}~\bibnamefont {Smerzi}},
  \bibinfo {author} {\bibfnamefont {M.~K.}\ \bibnamefont {Oberthaler}},
  \bibinfo {author} {\bibfnamefont {R.}~\bibnamefont {Schmied}},\ and\ \bibinfo
  {author} {\bibfnamefont {P.}~\bibnamefont {Treutlein}},\ }\bibfield  {title}
  {\bibinfo {title} {Quantum metrology with nonclassical states of atomic
  ensembles},\ }\href {https://doi.org/10.1103/RevModPhys.90.035005} {\bibfield
   {journal} {\bibinfo  {journal} {Rev. Mod. Phys.}\ }\textbf {\bibinfo
  {volume} {90}},\ \bibinfo {pages} {035005} (\bibinfo {year}
  {2018})}\BibitemShut {NoStop}%
\bibitem [{\citenamefont {Henderson}\ and\ \citenamefont
  {Vedral}(2001)}]{Vedral-2001}%
  \BibitemOpen
  \bibfield  {author} {\bibinfo {author} {\bibfnamefont {L.}~\bibnamefont
  {Henderson}}\ and\ \bibinfo {author} {\bibfnamefont {V.}~\bibnamefont
  {Vedral}},\ }\bibfield  {title} {\bibinfo {title} {Classical, quantum and
  total correlations},\ }\href {https://doi.org/10.1088/0305-4470/34/35/315}
  {\bibfield  {journal} {\bibinfo  {journal} {J. Phys. A: Math. Gen.}\ }\textbf
  {\bibinfo {volume} {34}},\ \bibinfo {pages} {6899} (\bibinfo {year}
  {2001})}\BibitemShut {NoStop}%
\bibitem [{\citenamefont {Ollivier}\ and\ \citenamefont
  {Zurek}(2001)}]{Zurek-2002}%
  \BibitemOpen
  \bibfield  {author} {\bibinfo {author} {\bibfnamefont {H.}~\bibnamefont
  {Ollivier}}\ and\ \bibinfo {author} {\bibfnamefont {W.~H.}\ \bibnamefont
  {Zurek}},\ }\bibfield  {title} {\bibinfo {title} {Quantum discord: A measure
  of the quantumness of correlations},\ }\href
  {https://doi.org/10.1103/PhysRevLett.88.017901} {\bibfield  {journal}
  {\bibinfo  {journal} {Phys. Rev. Lett.}\ }\textbf {\bibinfo {volume} {88}},\
  \bibinfo {pages} {017901} (\bibinfo {year} {2001})}\BibitemShut {NoStop}%
\bibitem [{\citenamefont {Werlang}\ and\ \citenamefont
  {Rigolin}(2010)}]{2010_ThermalDiscord}%
  \BibitemOpen
  \bibfield  {author} {\bibinfo {author} {\bibfnamefont {T.}~\bibnamefont
  {Werlang}}\ and\ \bibinfo {author} {\bibfnamefont {G.}~\bibnamefont
  {Rigolin}},\ }\bibfield  {title} {\bibinfo {title} {Thermal and magnetic
  quantum discord in $\mathrm{H}$eisenberg models},\ }\href
  {https://doi.org/10.1103/PhysRevA.81.044101} {\bibfield  {journal} {\bibinfo
  {journal} {Phys. Rev. A}\ }\textbf {\bibinfo {volume} {81}},\ \bibinfo
  {pages} {044101} (\bibinfo {year} {2010})}\BibitemShut {NoStop}%
\bibitem [{\citenamefont {Dillenschneider}(2008)}]{2008RaoulQDResource}%
  \BibitemOpen
  \bibfield  {author} {\bibinfo {author} {\bibfnamefont {R.}~\bibnamefont
  {Dillenschneider}},\ }\bibfield  {title} {\bibinfo {title} {Quantum discord
  and quantum phase transition in spin chains},\ }\href
  {https://doi.org/10.1103/PhysRevB.78.224413} {\bibfield  {journal} {\bibinfo
  {journal} {Phys. Rev. B}\ }\textbf {\bibinfo {volume} {78}},\ \bibinfo
  {pages} {224413} (\bibinfo {year} {2008})}\BibitemShut {NoStop}%
\bibitem [{\citenamefont {Daki\'{c}}\ \emph {et~al.}(2012)\citenamefont
  {Daki\'{c}}, \citenamefont {Lipp}, \citenamefont {Ma}, \citenamefont
  {Ringbauer}, \citenamefont {Kropatschek}, \citenamefont {Barz}, \citenamefont
  {Paterek}, \citenamefont {Vedral}, \citenamefont {Zeilinger}, \citenamefont
  {Brukner},\ and\ \citenamefont {Walther}}]{2012DakicRemoteSP}%
  \BibitemOpen
  \bibfield  {author} {\bibinfo {author} {\bibfnamefont {B.}~\bibnamefont
  {Daki\'{c}}}, \bibinfo {author} {\bibfnamefont {Y.~O.}\ \bibnamefont {Lipp}},
  \bibinfo {author} {\bibfnamefont {X.}~\bibnamefont {Ma}}, \bibinfo {author}
  {\bibfnamefont {M.}~\bibnamefont {Ringbauer}}, \bibinfo {author}
  {\bibfnamefont {S.}~\bibnamefont {Kropatschek}}, \bibinfo {author}
  {\bibfnamefont {S.}~\bibnamefont {Barz}}, \bibinfo {author} {\bibfnamefont
  {T.}~\bibnamefont {Paterek}}, \bibinfo {author} {\bibfnamefont
  {V.}~\bibnamefont {Vedral}}, \bibinfo {author} {\bibfnamefont
  {A.}~\bibnamefont {Zeilinger}}, \bibinfo {author} {\bibfnamefont
  {{\v{C}}.}~\bibnamefont {Brukner}},\ and\ \bibinfo {author} {\bibfnamefont
  {P.}~\bibnamefont {Walther}},\ }\bibfield  {title} {\bibinfo {title} {Quantum
  discord as resource for remote state preparation},\ }\href
  {https://doi.org/10.1038/nphys2377} {\bibfield  {journal} {\bibinfo
  {journal} {Nature Phys.}\ }\textbf {\bibinfo {volume} {8}},\ \bibinfo {pages}
  {666} (\bibinfo {year} {2012})}\BibitemShut {NoStop}%
\bibitem [{\citenamefont {Pirandola}(2014)}]{2014StefanoRDis}%
  \BibitemOpen
  \bibfield  {author} {\bibinfo {author} {\bibfnamefont {S.}~\bibnamefont
  {Pirandola}},\ }\bibfield  {title} {\bibinfo {title} {Quantum discord as a
  resource for quantum cryptography},\ }\href
  {https://doi.org/10.1038/srep06956} {\bibfield  {journal} {\bibinfo
  {journal} {Sci. Rep.}\ }\textbf {\bibinfo {volume} {4}},\ \bibinfo {pages}
  {6956} (\bibinfo {year} {2014})}\BibitemShut {NoStop}%
\bibitem [{\citenamefont {Liu}\ \emph {et~al.}(2020)\citenamefont {Liu},
  \citenamefont {Shang},\ and\ \citenamefont {Liu}}]{2020LiuDisResource}%
  \BibitemOpen
  \bibfield  {author} {\bibinfo {author} {\bibfnamefont {R.}~\bibnamefont
  {Liu}}, \bibinfo {author} {\bibfnamefont {T.}~\bibnamefont {Shang}},\ and\
  \bibinfo {author} {\bibfnamefont {J.-w.}\ \bibnamefont {Liu}},\ }\bibfield
  {title} {\bibinfo {title} {Quantum discord as a resource for quantum
  cryptography},\ }\href {https://doi.org/10.1007/s11128-019-2558-1} {\bibfield
   {journal} {\bibinfo  {journal} {Quantum Information Processing}\ }\textbf
  {\bibinfo {volume} {19}},\ \bibinfo {pages} {58} (\bibinfo {year}
  {2020})}\BibitemShut {NoStop}%
\bibitem [{\citenamefont {Madhok}\ and\ \citenamefont
  {Datta}(2013)}]{2013Datta}%
  \BibitemOpen
  \bibfield  {author} {\bibinfo {author} {\bibfnamefont {V.}~\bibnamefont
  {Madhok}}\ and\ \bibinfo {author} {\bibfnamefont {A.}~\bibnamefont {Datta}},\
  }\bibfield  {title} {\bibinfo {title} {Quantum discord as a resource in
  quantum communication},\ }\href {https://doi.org/10.1142/S0217979213450410}
  {\bibfield  {journal} {\bibinfo  {journal} {Inter. J. Mod. Phys. B}\ }\textbf
  {\bibinfo {volume} {27}},\ \bibinfo {pages} {1345041} (\bibinfo {year}
  {2013})}\BibitemShut {NoStop}%
\bibitem [{\citenamefont {Girolami}\ \emph {et~al.}(2013)\citenamefont
  {Girolami}, \citenamefont {Tufarelli},\ and\ \citenamefont
  {Adesso}}]{2013_Adesso_Metrology}%
  \BibitemOpen
  \bibfield  {author} {\bibinfo {author} {\bibfnamefont {D.}~\bibnamefont
  {Girolami}}, \bibinfo {author} {\bibfnamefont {T.}~\bibnamefont
  {Tufarelli}},\ and\ \bibinfo {author} {\bibfnamefont {G.}~\bibnamefont
  {Adesso}},\ }\bibfield  {title} {\bibinfo {title} {Characterizing
  nonclassical correlations via local quantum uncertainty},\ }\href
  {https://doi.org/10.1103/PhysRevLett.110.240402} {\bibfield  {journal}
  {\bibinfo  {journal} {Phys. Rev. Lett.}\ }\textbf {\bibinfo {volume} {110}},\
  \bibinfo {pages} {240402} (\bibinfo {year} {2013})}\BibitemShut {NoStop}%
\bibitem [{\citenamefont {Girolami}\ \emph {et~al.}(2014)\citenamefont
  {Girolami}, \citenamefont {Souza}, \citenamefont {Giovannetti}, \citenamefont
  {Tufarelli}, \citenamefont {Filgueiras}, \citenamefont {Sarthour},
  \citenamefont {Soares-Pinto}, \citenamefont {Oliveira},\ and\ \citenamefont
  {Adesso}}]{2014_Adesso_Metrology}%
  \BibitemOpen
  \bibfield  {author} {\bibinfo {author} {\bibfnamefont {D.}~\bibnamefont
  {Girolami}}, \bibinfo {author} {\bibfnamefont {A.~M.}\ \bibnamefont {Souza}},
  \bibinfo {author} {\bibfnamefont {V.}~\bibnamefont {Giovannetti}}, \bibinfo
  {author} {\bibfnamefont {T.}~\bibnamefont {Tufarelli}}, \bibinfo {author}
  {\bibfnamefont {J.~G.}\ \bibnamefont {Filgueiras}}, \bibinfo {author}
  {\bibfnamefont {R.~S.}\ \bibnamefont {Sarthour}}, \bibinfo {author}
  {\bibfnamefont {D.~O.}\ \bibnamefont {Soares-Pinto}}, \bibinfo {author}
  {\bibfnamefont {I.~S.}\ \bibnamefont {Oliveira}},\ and\ \bibinfo {author}
  {\bibfnamefont {G.}~\bibnamefont {Adesso}},\ }\bibfield  {title} {\bibinfo
  {title} {Quantum discord determines the interferometric power of quantum
  states},\ }\href {https://doi.org/10.1103/PhysRevLett.112.210401} {\bibfield
  {journal} {\bibinfo  {journal} {Phys. Rev. Lett.}\ }\textbf {\bibinfo
  {volume} {112}},\ \bibinfo {pages} {210401} (\bibinfo {year}
  {2014})}\BibitemShut {NoStop}%
\bibitem [{\citenamefont {Sone}\ \emph {et~al.}(2019)\citenamefont {Sone},
  \citenamefont {Zhuang}, \citenamefont {Li}, \citenamefont {Liu},\ and\
  \citenamefont {Cappellaro}}]{2019_Metrology}%
  \BibitemOpen
  \bibfield  {author} {\bibinfo {author} {\bibfnamefont {A.}~\bibnamefont
  {Sone}}, \bibinfo {author} {\bibfnamefont {Q.}~\bibnamefont {Zhuang}},
  \bibinfo {author} {\bibfnamefont {C.}~\bibnamefont {Li}}, \bibinfo {author}
  {\bibfnamefont {Y.-X.}\ \bibnamefont {Liu}},\ and\ \bibinfo {author}
  {\bibfnamefont {P.}~\bibnamefont {Cappellaro}},\ }\bibfield  {title}
  {\bibinfo {title} {Nonclassical correlations for quantum metrology in thermal
  equilibrium},\ }\href {https://doi.org/10.1103/PhysRevA.99.052318} {\bibfield
   {journal} {\bibinfo  {journal} {Phys. Rev. A}\ }\textbf {\bibinfo {volume}
  {99}},\ \bibinfo {pages} {052318} (\bibinfo {year} {2019})}\BibitemShut
  {NoStop}%
\bibitem [{\citenamefont {Micadei}\ \emph {et~al.}(2019)\citenamefont
  {Micadei}, \citenamefont {Peterson}, \citenamefont {Souza},\ and\
  \citenamefont {et~al.}}]{AHF_2019_NatCommun_Lutz}%
  \BibitemOpen
  \bibfield  {author} {\bibinfo {author} {\bibfnamefont {K.}~\bibnamefont
  {Micadei}}, \bibinfo {author} {\bibfnamefont {J.~P.~S.}\ \bibnamefont
  {Peterson}}, \bibinfo {author} {\bibfnamefont {A.~M.}\ \bibnamefont
  {Souza}},\ and\ \bibinfo {author} {\bibnamefont {et~al.}},\ }\bibfield
  {title} {\bibinfo {title} {Reversing the direction of heat flow using quantum
  correlations},\ }\href {https://doi.org/10.1038/s41467-019-10333-7}
  {\bibfield  {journal} {\bibinfo  {journal} {Nat. Commun.}\ }\textbf {\bibinfo
  {volume} {10}},\ \bibinfo {pages} {2456} (\bibinfo {year}
  {2019})}\BibitemShut {NoStop}%
\bibitem [{\citenamefont {Pusuluk}\ and\ \citenamefont
  {M\"{u}stecapl{\i}o\u{g}lu}(2021)}]{Pusuluk_2021_PRR}%
  \BibitemOpen
  \bibfield  {author} {\bibinfo {author} {\bibfnamefont {O.}~\bibnamefont
  {Pusuluk}}\ and\ \bibinfo {author} {\bibfnamefont {O.~E.}\ \bibnamefont
  {M\"{u}stecapl{\i}o\u{g}lu}},\ }\bibfield  {title} {\bibinfo {title} {Quantum
  $\mathrm{R}$ayleigh problem and thermocoherent $\mathrm{O}$nsager
  relations},\ }\href {https://doi.org/10.1103/PhysRevResearch.3.023235}
  {\bibfield  {journal} {\bibinfo  {journal} {Phys. Rev. Research}\ }\textbf
  {\bibinfo {volume} {3}},\ \bibinfo {pages} {023235} (\bibinfo {year}
  {2021})}\BibitemShut {NoStop}%
\bibitem [{\citenamefont {White}\ and\ \citenamefont
  {Martin}(1999)}]{1999_qChemDMRG}%
  \BibitemOpen
  \bibfield  {author} {\bibinfo {author} {\bibfnamefont {S.~R.}\ \bibnamefont
  {White}}\ and\ \bibinfo {author} {\bibfnamefont {R.~L.}\ \bibnamefont
  {Martin}},\ }\bibfield  {title} {\bibinfo {title} {Ab initio quantum
  chemistry using the density matrix renormalization group},\ }\href
  {https://doi.org/10.1063/1.478295} {\bibfield  {journal} {\bibinfo  {journal}
  {The J. of Chem. Phys.}\ }\textbf {\bibinfo {volume} {110}},\ \bibinfo
  {pages} {4127} (\bibinfo {year} {1999})}\BibitemShut {NoStop}%
\bibitem [{\citenamefont {Daul}\ \emph {et~al.}(2000)\citenamefont {Daul},
  \citenamefont {Ciofini}, \citenamefont {Daul},\ and\ \citenamefont
  {White}}]{2000_qChemDMRG}%
  \BibitemOpen
  \bibfield  {author} {\bibinfo {author} {\bibfnamefont {S.}~\bibnamefont
  {Daul}}, \bibinfo {author} {\bibfnamefont {I.}~\bibnamefont {Ciofini}},
  \bibinfo {author} {\bibfnamefont {C.}~\bibnamefont {Daul}},\ and\ \bibinfo
  {author} {\bibfnamefont {S.~R.}\ \bibnamefont {White}},\ }\bibfield  {title}
  {\bibinfo {title} {Full-$\mathrm{CI}$ quantum chemistry using the density
  matrix renormalization group},\ }\href
  {https://doi.org/10.1002/1097-461X(2000)79:6<331::AID-QUA1>3.0.CO;2-Y}
  {\bibfield  {journal} {\bibinfo  {journal} {Int. J. Quantum Chem.}\ }\textbf
  {\bibinfo {volume} {79}},\ \bibinfo {pages} {331} (\bibinfo {year}
  {2000})}\BibitemShut {NoStop}%
\bibitem [{\citenamefont {Legeza}\ and\ \citenamefont
  {S\'olyom}(2003)}]{2003_qChemDMRG_MutInfo}%
  \BibitemOpen
  \bibfield  {author} {\bibinfo {author} {\bibfnamefont {O.}~\bibnamefont
  {Legeza}}\ and\ \bibinfo {author} {\bibfnamefont {J.}~\bibnamefont
  {S\'olyom}},\ }\bibfield  {title} {\bibinfo {title} {Optimizing the
  density-matrix renormalization group method using quantum information
  entropy},\ }\href {https://doi.org/10.1103/PhysRevB.68.195116} {\bibfield
  {journal} {\bibinfo  {journal} {Phys. Rev. B}\ }\textbf {\bibinfo {volume}
  {68}},\ \bibinfo {pages} {195116} (\bibinfo {year} {2003})}\BibitemShut
  {NoStop}%
\bibitem [{\citenamefont {Rissler}\ \emph {et~al.}(2006)\citenamefont
  {Rissler}, \citenamefont {Noack},\ and\ \citenamefont
  {White}}]{2006_qChemDMRG_MutInfo}%
  \BibitemOpen
  \bibfield  {author} {\bibinfo {author} {\bibfnamefont {J.}~\bibnamefont
  {Rissler}}, \bibinfo {author} {\bibfnamefont {R.}~\bibnamefont {Noack}},\
  and\ \bibinfo {author} {\bibfnamefont {S.}~\bibnamefont {White}},\ }\bibfield
   {title} {\bibinfo {title} {Measuring orbital interaction using quantum
  information theory},\ }\href {https://doi.org/10.1016/j.chemphys.2005.10.018}
  {\bibfield  {journal} {\bibinfo  {journal} {Chem. Phys.}\ }\textbf {\bibinfo
  {volume} {323}},\ \bibinfo {pages} {519} (\bibinfo {year}
  {2006})}\BibitemShut {NoStop}%
\bibitem [{\citenamefont {Szalay}\ \emph {et~al.}(2015)\citenamefont {Szalay},
  \citenamefont {Pfeffer}, \citenamefont {Murg}, \citenamefont {Barcza},
  \citenamefont {Verstraete}, \citenamefont {Schneider},\ and\ \citenamefont
  {Legeza}}]{2015_qChemDMRG_MutInfo}%
  \BibitemOpen
  \bibfield  {author} {\bibinfo {author} {\bibfnamefont {S.}~\bibnamefont
  {Szalay}}, \bibinfo {author} {\bibfnamefont {M.}~\bibnamefont {Pfeffer}},
  \bibinfo {author} {\bibfnamefont {V.}~\bibnamefont {Murg}}, \bibinfo {author}
  {\bibfnamefont {G.}~\bibnamefont {Barcza}}, \bibinfo {author} {\bibfnamefont
  {F.}~\bibnamefont {Verstraete}}, \bibinfo {author} {\bibfnamefont
  {R.}~\bibnamefont {Schneider}},\ and\ \bibinfo {author} {\bibfnamefont
  {O.}~\bibnamefont {Legeza}},\ }\bibfield  {title} {\bibinfo {title} {Tensor
  product methods and entanglement optimization for ab initio quantum
  chemistry},\ }\href {https://doi.org/10.1002/qua.24898} {\bibfield  {journal}
  {\bibinfo  {journal} {Int. J. Quantum Chem.}\ }\textbf {\bibinfo {volume}
  {115}},\ \bibinfo {pages} {1342} (\bibinfo {year} {2015})}\BibitemShut
  {NoStop}%
\bibitem [{\citenamefont {Stein}\ and\ \citenamefont
  {Reiher}(2016)}]{2016_qChemDMRG_MutInfo}%
  \BibitemOpen
  \bibfield  {author} {\bibinfo {author} {\bibfnamefont {C.~J.}\ \bibnamefont
  {Stein}}\ and\ \bibinfo {author} {\bibfnamefont {M.}~\bibnamefont {Reiher}},\
  }\bibfield  {title} {\bibinfo {title} {Automated selection of active orbital
  spaces},\ }\href {https://doi.org/10.1021/acs.jctc.6b00156} {\bibfield
  {journal} {\bibinfo  {journal} {J. Chem. Theory Comput.}\ }\textbf {\bibinfo
  {volume} {12}},\ \bibinfo {pages} {1760} (\bibinfo {year}
  {2016})}\BibitemShut {NoStop}%
\bibitem [{\citenamefont {Krumnow}\ \emph {et~al.}(2016)\citenamefont
  {Krumnow}, \citenamefont {Veis}, \citenamefont {Legeza},\ and\ \citenamefont
  {Eisert}}]{2016_qChemDMRG_MutInfo_2}%
  \BibitemOpen
  \bibfield  {author} {\bibinfo {author} {\bibfnamefont {C.}~\bibnamefont
  {Krumnow}}, \bibinfo {author} {\bibfnamefont {L.}~\bibnamefont {Veis}},
  \bibinfo {author} {\bibfnamefont {O.}~\bibnamefont {Legeza}},\ and\ \bibinfo
  {author} {\bibfnamefont {J.}~\bibnamefont {Eisert}},\ }\bibfield  {title}
  {\bibinfo {title} {Fermionic orbital optimization in tensor network states},\
  }\href {https://doi.org/10.1103/PhysRevLett.117.210402} {\bibfield  {journal}
  {\bibinfo  {journal} {Phys. Rev. Lett.}\ }\textbf {\bibinfo {volume} {117}},\
  \bibinfo {pages} {210402} (\bibinfo {year} {2016})}\BibitemShut {NoStop}%
\bibitem [{\citenamefont {Barcza}\ \emph {et~al.}(2011)\citenamefont {Barcza},
  \citenamefont {Legeza}, \citenamefont {Marti},\ and\ \citenamefont
  {Reiher}}]{2011_ChemBond_MutInfo}%
  \BibitemOpen
  \bibfield  {author} {\bibinfo {author} {\bibfnamefont {G.}~\bibnamefont
  {Barcza}}, \bibinfo {author} {\bibfnamefont {O.}~\bibnamefont {Legeza}},
  \bibinfo {author} {\bibfnamefont {K.~H.}\ \bibnamefont {Marti}},\ and\
  \bibinfo {author} {\bibfnamefont {M.}~\bibnamefont {Reiher}},\ }\bibfield
  {title} {\bibinfo {title} {Quantum-information analysis of electronic states
  of different molecular structures},\ }\href
  {https://doi.org/10.1103/PhysRevA.83.012508} {\bibfield  {journal} {\bibinfo
  {journal} {Phys. Rev. A}\ }\textbf {\bibinfo {volume} {83}},\ \bibinfo
  {pages} {012508} (\bibinfo {year} {2011})}\BibitemShut {NoStop}%
\bibitem [{\citenamefont {Boguslawski}\ \emph {et~al.}(2012)\citenamefont
  {Boguslawski}, \citenamefont {Tecmer}, \citenamefont {Legeza},\ and\
  \citenamefont {Reiher}}]{2012_ChemBond_MutInfo}%
  \BibitemOpen
  \bibfield  {author} {\bibinfo {author} {\bibfnamefont {K.}~\bibnamefont
  {Boguslawski}}, \bibinfo {author} {\bibfnamefont {P.}~\bibnamefont {Tecmer}},
  \bibinfo {author} {\bibfnamefont {O.}~\bibnamefont {Legeza}},\ and\ \bibinfo
  {author} {\bibfnamefont {M.}~\bibnamefont {Reiher}},\ }\bibfield  {title}
  {\bibinfo {title} {Entanglement measures for single- and multireference
  correlation effects},\ }\href {https://doi.org/10.1021/jz301319v} {\bibfield
  {journal} {\bibinfo  {journal} {J. Phys. Chem. Lett.}\ }\textbf {\bibinfo
  {volume} {3}},\ \bibinfo {pages} {3129} (\bibinfo {year} {2012})}\BibitemShut
  {NoStop}%
\bibitem [{\citenamefont {Boguslawski}\ \emph {et~al.}(2013)\citenamefont
  {Boguslawski}, \citenamefont {Tecmer}, \citenamefont {Barcza}, \citenamefont
  {Legeza},\ and\ \citenamefont {Reiher}}]{2013_ChemBond_MutInfo}%
  \BibitemOpen
  \bibfield  {author} {\bibinfo {author} {\bibfnamefont {K.}~\bibnamefont
  {Boguslawski}}, \bibinfo {author} {\bibfnamefont {P.}~\bibnamefont {Tecmer}},
  \bibinfo {author} {\bibfnamefont {G.}~\bibnamefont {Barcza}}, \bibinfo
  {author} {\bibfnamefont {O.}~\bibnamefont {Legeza}},\ and\ \bibinfo {author}
  {\bibfnamefont {M.}~\bibnamefont {Reiher}},\ }\bibfield  {title} {\bibinfo
  {title} {Orbital entanglement in bond-formation processes},\ }\href
  {https://doi.org/10.1021/ct400247p} {\bibfield  {journal} {\bibinfo
  {journal} {J. Chem. Theory Comput.}\ }\textbf {\bibinfo {volume} {9}},\
  \bibinfo {pages} {2959} (\bibinfo {year} {2013})}\BibitemShut {NoStop}%
\bibitem [{\citenamefont {Kurashige}\ \emph {et~al.}(2013)\citenamefont
  {Kurashige}, \citenamefont {Chan},\ and\ \citenamefont
  {Yanai}}]{2013_ChemBond_MutInfo_2}%
  \BibitemOpen
  \bibfield  {author} {\bibinfo {author} {\bibfnamefont {Y.}~\bibnamefont
  {Kurashige}}, \bibinfo {author} {\bibfnamefont {G.}~\bibnamefont {Chan}},\
  and\ \bibinfo {author} {\bibfnamefont {T.}~\bibnamefont {Yanai}},\ }\bibfield
   {title} {\bibinfo {title} {Entangled quantum electronic wavefunctions of the
  $\mathrm{Mn}_4\mathrm{CaO}_5$ cluster in photosystem $\mathrm{II}$},\ }\href
  {https://doi.org/10.1038/nchem.1677} {\bibfield  {journal} {\bibinfo
  {journal} {Nature Chem.}\ }\textbf {\bibinfo {volume} {5}},\ \bibinfo {pages}
  {660} (\bibinfo {year} {2013})}\BibitemShut {NoStop}%
\bibitem [{\citenamefont {Mottet}\ \emph {et~al.}(2014)\citenamefont {Mottet},
  \citenamefont {Tecmer}, \citenamefont {Boguslawski}, , \citenamefont
  {Legeza},\ and\ \citenamefont {Reiher}}]{2014_ChemBond_MutInfo}%
  \BibitemOpen
  \bibfield  {author} {\bibinfo {author} {\bibfnamefont {M.}~\bibnamefont
  {Mottet}}, \bibinfo {author} {\bibfnamefont {P.}~\bibnamefont {Tecmer}},
  \bibinfo {author} {\bibfnamefont {K.}~\bibnamefont {Boguslawski}}, , \bibinfo
  {author} {\bibfnamefont {O.}~\bibnamefont {Legeza}},\ and\ \bibinfo {author}
  {\bibfnamefont {M.}~\bibnamefont {Reiher}},\ }\bibfield  {title} {\bibinfo
  {title} {Quantum entanglement in carbon-carbon, carbon-phosphorus and
  silicon-silicon bonds},\ }\href {https://doi.org/10.1039/C4CP00277F}
  {\bibfield  {journal} {\bibinfo  {journal} {Phys. Chem. Chem. Phys.}\
  }\textbf {\bibinfo {volume} {16}},\ \bibinfo {pages} {8872} (\bibinfo {year}
  {2014})}\BibitemShut {NoStop}%
\bibitem [{\citenamefont {Szalay}\ \emph {et~al.}(2017)\citenamefont {Szalay},
  \citenamefont {Barcza}, \citenamefont {Szilv\'{a}si},\ and\ \citenamefont
  {Legeza}}]{2017_ChemBond_MutInfo}%
  \BibitemOpen
  \bibfield  {author} {\bibinfo {author} {\bibfnamefont {S.}~\bibnamefont
  {Szalay}}, \bibinfo {author} {\bibfnamefont {G.}~\bibnamefont {Barcza}},
  \bibinfo {author} {\bibfnamefont {T.}~\bibnamefont {Szilv\'{a}si}},\ and\
  \bibinfo {author} {\bibfnamefont {O.}~\bibnamefont {Legeza}},\ }\bibfield
  {title} {\bibinfo {title} {The correlation theory of the chemical bond},\
  }\href {https://doi.org/10.1038/s41598-017-02447-z} {\bibfield  {journal}
  {\bibinfo  {journal} {Sci. Rep.}\ }\textbf {\bibinfo {volume} {7}},\ \bibinfo
  {pages} {2237} (\bibinfo {year} {2017})}\BibitemShut {NoStop}%
\bibitem [{\citenamefont {Stemmle}\ \emph {et~al.}(2018)\citenamefont
  {Stemmle}, \citenamefont {Paulus},\ and\ \citenamefont
  {Legeza}}]{2018_ChemBond_MutInfo}%
  \BibitemOpen
  \bibfield  {author} {\bibinfo {author} {\bibfnamefont {C.}~\bibnamefont
  {Stemmle}}, \bibinfo {author} {\bibfnamefont {B.}~\bibnamefont {Paulus}},\
  and\ \bibinfo {author} {\bibfnamefont {O.}~\bibnamefont {Legeza}},\
  }\bibfield  {title} {\bibinfo {title} {Analysis of electron-correlation
  effects in strongly correlated systems ($\mathrm{N}_{2}$ and
  $\mathrm{N}_{2}^{+}$) by applying the density-matrix renormalization-group
  method and quantum information theory},\ }\href
  {https://doi.org/10.1103/PhysRevA.97.022505} {\bibfield  {journal} {\bibinfo
  {journal} {Phys. Rev. A}\ }\textbf {\bibinfo {volume} {97}},\ \bibinfo
  {pages} {022505} (\bibinfo {year} {2018})}\BibitemShut {NoStop}%
\bibitem [{\citenamefont {Ding}\ \emph {et~al.}(2021)\citenamefont {Ding},
  \citenamefont {Mardazad}, \citenamefont {Das}, \citenamefont {Szalay},
  \citenamefont {Schollw\"{o}ck}, \citenamefont {Zimbor\'{a}s},\ and\
  \citenamefont {Schilling}}]{2021_JCTC_DMRG_SSR}%
  \BibitemOpen
  \bibfield  {author} {\bibinfo {author} {\bibfnamefont {L.}~\bibnamefont
  {Ding}}, \bibinfo {author} {\bibfnamefont {S.}~\bibnamefont {Mardazad}},
  \bibinfo {author} {\bibfnamefont {S.}~\bibnamefont {Das}}, \bibinfo {author}
  {\bibfnamefont {S.}~\bibnamefont {Szalay}}, \bibinfo {author} {\bibfnamefont
  {U.}~\bibnamefont {Schollw\"{o}ck}}, \bibinfo {author} {\bibfnamefont
  {Z.}~\bibnamefont {Zimbor\'{a}s}},\ and\ \bibinfo {author} {\bibfnamefont
  {C.}~\bibnamefont {Schilling}},\ }\bibfield  {title} {\bibinfo {title}
  {Concept of orbital entanglement and correlation in quantum chemistry},\
  }\href {https://doi.org/10.1021/acs.jctc.0c00559} {\bibfield  {journal}
  {\bibinfo  {journal} {J. Chem. Theory Comput.}\ }\textbf {\bibinfo {volume}
  {17}},\ \bibinfo {pages} {79} (\bibinfo {year} {2021})}\BibitemShut {NoStop}%
\bibitem [{\citenamefont {Modi}\ \emph {et~al.}(2010)\citenamefont {Modi},
  \citenamefont {Paterek}, \citenamefont {Son}, \citenamefont {Vedral},\ and\
  \citenamefont {Williamson}}]{2010_PRL_VV_CorrWithRelEnt}%
  \BibitemOpen
  \bibfield  {author} {\bibinfo {author} {\bibfnamefont {K.}~\bibnamefont
  {Modi}}, \bibinfo {author} {\bibfnamefont {T.}~\bibnamefont {Paterek}},
  \bibinfo {author} {\bibfnamefont {W.}~\bibnamefont {Son}}, \bibinfo {author}
  {\bibfnamefont {V.}~\bibnamefont {Vedral}},\ and\ \bibinfo {author}
  {\bibfnamefont {M.}~\bibnamefont {Williamson}},\ }\bibfield  {title}
  {\bibinfo {title} {Unified view of quantum and classical correlations},\
  }\href {https://doi.org/10.1103/PhysRevLett.104.080501} {\bibfield  {journal}
  {\bibinfo  {journal} {Phys. Rev. Lett.}\ }\textbf {\bibinfo {volume} {104}},\
  \bibinfo {pages} {080501} (\bibinfo {year} {2010})}\BibitemShut {NoStop}%
\bibitem [{\citenamefont {Shapourian}\ \emph
  {et~al.}(2017{\natexlab{a}})\citenamefont {Shapourian}, \citenamefont
  {Shiozaki},\ and\ \citenamefont {Ryu}}]{2017_PRB_PTandNegativity}%
  \BibitemOpen
  \bibfield  {author} {\bibinfo {author} {\bibfnamefont {H.}~\bibnamefont
  {Shapourian}}, \bibinfo {author} {\bibfnamefont {K.}~\bibnamefont
  {Shiozaki}},\ and\ \bibinfo {author} {\bibfnamefont {S.}~\bibnamefont
  {Ryu}},\ }\bibfield  {title} {\bibinfo {title} {Partial time-reversal
  transformation and entanglement negativity in fermionic systems},\ }\href
  {https://doi.org/10.1103/PhysRevB.95.165101} {\bibfield  {journal} {\bibinfo
  {journal} {Phys. Rev. B}\ }\textbf {\bibinfo {volume} {95}},\ \bibinfo
  {pages} {165101} (\bibinfo {year} {2017}{\natexlab{a}})}\BibitemShut
  {NoStop}%
\bibitem [{\citenamefont {Shapourian}\ \emph
  {et~al.}(2017{\natexlab{b}})\citenamefont {Shapourian}, \citenamefont
  {Shiozaki},\ and\ \citenamefont {Ryu}}]{2017_PRL_PTandNegativity}%
  \BibitemOpen
  \bibfield  {author} {\bibinfo {author} {\bibfnamefont {H.}~\bibnamefont
  {Shapourian}}, \bibinfo {author} {\bibfnamefont {K.}~\bibnamefont
  {Shiozaki}},\ and\ \bibinfo {author} {\bibfnamefont {S.}~\bibnamefont
  {Ryu}},\ }\bibfield  {title} {\bibinfo {title} {Many-body topological
  invariants for fermionic symmetry-protected topological phases},\ }\href
  {https://doi.org/10.1103/PhysRevLett.118.216402} {\bibfield  {journal}
  {\bibinfo  {journal} {Phys. Rev. Lett.}\ }\textbf {\bibinfo {volume} {118}},\
  \bibinfo {pages} {216402} (\bibinfo {year} {2017}{\natexlab{b}})}\BibitemShut
  {NoStop}%
\bibitem [{\citenamefont {Shiozaki}\ \emph {et~al.}(2018)\citenamefont
  {Shiozaki}, \citenamefont {Shapourian}, \citenamefont {Gomi},\ and\
  \citenamefont {Ryu}}]{2018_PRB_PTandNegativity}%
  \BibitemOpen
  \bibfield  {author} {\bibinfo {author} {\bibfnamefont {K.}~\bibnamefont
  {Shiozaki}}, \bibinfo {author} {\bibfnamefont {H.}~\bibnamefont
  {Shapourian}}, \bibinfo {author} {\bibfnamefont {K.}~\bibnamefont {Gomi}},\
  and\ \bibinfo {author} {\bibfnamefont {S.}~\bibnamefont {Ryu}},\ }\bibfield
  {title} {\bibinfo {title} {Many-body topological invariants for fermionic
  short-range entangled topological phases protected by antiunitary
  symmetries},\ }\href {https://doi.org/10.1103/PhysRevB.98.035151} {\bibfield
  {journal} {\bibinfo  {journal} {Phys. Rev. B}\ }\textbf {\bibinfo {volume}
  {98}},\ \bibinfo {pages} {035151} (\bibinfo {year} {2018})}\BibitemShut
  {NoStop}%
\bibitem [{\citenamefont {Shapourian}\ and\ \citenamefont
  {Ryu}(2019)}]{2019_PRA_PTandNegativity}%
  \BibitemOpen
  \bibfield  {author} {\bibinfo {author} {\bibfnamefont {H.}~\bibnamefont
  {Shapourian}}\ and\ \bibinfo {author} {\bibfnamefont {S.}~\bibnamefont
  {Ryu}},\ }\bibfield  {title} {\bibinfo {title} {Entanglement negativity of
  fermions: Monotonicity, separability criterion, and classification of
  few-mode states},\ }\href {https://doi.org/10.1103/PhysRevA.99.022310}
  {\bibfield  {journal} {\bibinfo  {journal} {Phys. Rev. A}\ }\textbf {\bibinfo
  {volume} {99}},\ \bibinfo {pages} {022310} (\bibinfo {year}
  {2019})}\BibitemShut {NoStop}%
\bibitem [{\citenamefont {Pople}\ \emph {et~al.}(1977)\citenamefont {Pople},
  \citenamefont {Seeger},\ and\ \citenamefont {Krishnan}}]{CI_77}%
  \BibitemOpen
  \bibfield  {author} {\bibinfo {author} {\bibfnamefont {J.~A.}\ \bibnamefont
  {Pople}}, \bibinfo {author} {\bibfnamefont {R.}~\bibnamefont {Seeger}},\ and\
  \bibinfo {author} {\bibfnamefont {R.}~\bibnamefont {Krishnan}},\ }\bibfield
  {title} {\bibinfo {title} {Variational configuration interaction methods and
  comparison with perturbation theory},\ }\href
  {https://doi.org/https://doi.org/10.1002/qua.560120820} {\bibfield  {journal}
  {\bibinfo  {journal} {Int. J. Quantum Chem.}\ }\textbf {\bibinfo {volume}
  {12}},\ \bibinfo {pages} {149} (\bibinfo {year} {1977})}\BibitemShut
  {NoStop}%
\bibitem [{\citenamefont {Krishnan}\ \emph {et~al.}(1980)\citenamefont
  {Krishnan}, \citenamefont {Schlegel},\ and\ \citenamefont {Pople}}]{CI_80}%
  \BibitemOpen
  \bibfield  {author} {\bibinfo {author} {\bibfnamefont {R.}~\bibnamefont
  {Krishnan}}, \bibinfo {author} {\bibfnamefont {H.~B.}\ \bibnamefont
  {Schlegel}},\ and\ \bibinfo {author} {\bibfnamefont {J.~A.}\ \bibnamefont
  {Pople}},\ }\bibfield  {title} {\bibinfo {title} {Derivative studies in
  configuration-interaction theory},\ }\href {https://doi.org/10.1063/1.439708}
  {\bibfield  {journal} {\bibinfo  {journal} {J. Chem. Phys.}\ }\textbf
  {\bibinfo {volume} {72}},\ \bibinfo {pages} {4654} (\bibinfo {year}
  {1980})}\BibitemShut {NoStop}%
\bibitem [{\citenamefont {Raghavachari}\ and\ \citenamefont
  {Pople}(1981)}]{CI_81}%
  \BibitemOpen
  \bibfield  {author} {\bibinfo {author} {\bibfnamefont {K.}~\bibnamefont
  {Raghavachari}}\ and\ \bibinfo {author} {\bibfnamefont {J.~A.}\ \bibnamefont
  {Pople}},\ }\bibfield  {title} {\bibinfo {title} {Calculation of one-electron
  properties using limited configuration interaction techniques},\ }\href
  {https://doi.org/https://doi.org/10.1002/qua.560200503} {\bibfield  {journal}
  {\bibinfo  {journal} {Int. J. Quantum Chem.}\ }\textbf {\bibinfo {volume}
  {20}},\ \bibinfo {pages} {1067} (\bibinfo {year} {1981})}\BibitemShut
  {NoStop}%
\bibitem [{\citenamefont {Frisch}\ and\ \citenamefont {et~al.}()}]{g09}%
  \BibitemOpen
  \bibfield  {author} {\bibinfo {author} {\bibfnamefont {M.}~\bibnamefont
  {Frisch}}\ and\ \bibinfo {author} {\bibnamefont {et~al.}},\ }\href@noop {}
  {\bibinfo {title} {Gaussian~09 {R}evision {E}.01}},\ \bibinfo {note}
  {gaussian Inc. Wallingford CT 2009}\BibitemShut {NoStop}%
\bibitem [{\citenamefont {Friis}\ \emph {et~al.}(2013)\citenamefont {Friis},
  \citenamefont {Lee},\ and\ \citenamefont {Bruschi}}]{2013_PRA_fPTrace}%
  \BibitemOpen
  \bibfield  {author} {\bibinfo {author} {\bibfnamefont {N.}~\bibnamefont
  {Friis}}, \bibinfo {author} {\bibfnamefont {A.~R.}\ \bibnamefont {Lee}},\
  and\ \bibinfo {author} {\bibfnamefont {D.~E.}\ \bibnamefont {Bruschi}},\
  }\bibfield  {title} {\bibinfo {title} {Fermionic-mode entanglement in quantum
  information},\ }\href {https://doi.org/10.1103/PhysRevA.87.022338} {\bibfield
   {journal} {\bibinfo  {journal} {Phys. Rev. A}\ }\textbf {\bibinfo {volume}
  {87}},\ \bibinfo {pages} {022338} (\bibinfo {year} {2013})}\BibitemShut
  {NoStop}%
\bibitem [{\citenamefont {Vidal}\ \emph {et~al.}(2021)\citenamefont {Vidal},
  \citenamefont {Bera}, \citenamefont {Riera}, \citenamefont {Lewenstein},\
  and\ \citenamefont {Bera}}]{2021_arXiv_fPTrace}%
  \BibitemOpen
  \bibfield  {author} {\bibinfo {author} {\bibfnamefont {N.~T.}\ \bibnamefont
  {Vidal}}, \bibinfo {author} {\bibfnamefont {M.~L.}\ \bibnamefont {Bera}},
  \bibinfo {author} {\bibfnamefont {A.}~\bibnamefont {Riera}}, \bibinfo
  {author} {\bibfnamefont {M.}~\bibnamefont {Lewenstein}},\ and\ \bibinfo
  {author} {\bibfnamefont {M.~N.}\ \bibnamefont {Bera}},\ }\bibfield  {title}
  {\bibinfo {title} {Quantum operations in an information theory for
  fermions},\ }\href {https://arxiv.org/abs/2102.09074} {\bibfield  {journal}
  {\bibinfo  {journal} {arXiv:2102.09074 [quant-ph]}\ } (\bibinfo {year}
  {2021})}\BibitemShut {NoStop}%
\bibitem [{\citenamefont {Wiseman}\ and\ \citenamefont
  {Vaccaro}(2003)}]{FixedN_2003_PRA}%
  \BibitemOpen
  \bibfield  {author} {\bibinfo {author} {\bibfnamefont {H.~M.}\ \bibnamefont
  {Wiseman}}\ and\ \bibinfo {author} {\bibfnamefont {J.~A.}\ \bibnamefont
  {Vaccaro}},\ }\bibfield  {title} {\bibinfo {title} {Entanglement of
  indistinguishable particles shared between two parties},\ }\href
  {https://doi.org/10.1103/PhysRevLett.91.097902} {\bibfield  {journal}
  {\bibinfo  {journal} {Phys. Rev. Lett.}\ }\textbf {\bibinfo {volume} {91}},\
  \bibinfo {pages} {097902} (\bibinfo {year} {2003})}\BibitemShut {NoStop}%
\bibitem [{\citenamefont {Montero}\ and\ \citenamefont
  {Mart\'{\i}n-Mart\'{\i}nez}(2011)}]{PhysRevA.83.062323}%
  \BibitemOpen
  \bibfield  {author} {\bibinfo {author} {\bibfnamefont {M.}~\bibnamefont
  {Montero}}\ and\ \bibinfo {author} {\bibfnamefont {E.}~\bibnamefont
  {Mart\'{\i}n-Mart\'{\i}nez}},\ }\bibfield  {title} {\bibinfo {title}
  {Fermionic entanglement ambiguity in noninertial frames},\ }\href
  {https://doi.org/10.1103/PhysRevA.83.062323} {\bibfield  {journal} {\bibinfo
  {journal} {Phys. Rev. A}\ }\textbf {\bibinfo {volume} {83}},\ \bibinfo
  {pages} {062323} (\bibinfo {year} {2011})}\BibitemShut {NoStop}%
\bibitem [{\citenamefont {Johansson}(2016)}]{arXiv_1610.00539}%
  \BibitemOpen
  \bibfield  {author} {\bibinfo {author} {\bibfnamefont {M.}~\bibnamefont
  {Johansson}},\ }\bibfield  {title} {\bibinfo {title} {Comment on `reasonable
  fermionic quantum information theories require relativity'},\ }\href
  {https://arxiv.org/abs/1610.00539} {\bibfield  {journal} {\bibinfo  {journal}
  {arXiv:1610.00539 [quant-ph]}\ } (\bibinfo {year} {2016})}\BibitemShut
  {NoStop}%
\bibitem [{\citenamefont {Wick}\ \emph {et~al.}(1952)\citenamefont {Wick},
  \citenamefont {Wightman},\ and\ \citenamefont {Wigner}}]{PhysRev.88.101}%
  \BibitemOpen
  \bibfield  {author} {\bibinfo {author} {\bibfnamefont {G.~C.}\ \bibnamefont
  {Wick}}, \bibinfo {author} {\bibfnamefont {A.~S.}\ \bibnamefont {Wightman}},\
  and\ \bibinfo {author} {\bibfnamefont {E.~P.}\ \bibnamefont {Wigner}},\
  }\bibfield  {title} {\bibinfo {title} {The intrinsic parity of elementary
  particles},\ }\href {https://doi.org/10.1103/PhysRev.88.101} {\bibfield
  {journal} {\bibinfo  {journal} {Phys. Rev.}\ }\textbf {\bibinfo {volume}
  {88}},\ \bibinfo {pages} {101} (\bibinfo {year} {1952})}\BibitemShut
  {NoStop}%
\bibitem [{\citenamefont {Jungnitsch}\ \emph {et~al.}(2011)\citenamefont
  {Jungnitsch}, \citenamefont {Moroder},\ and\ \citenamefont
  {G\"{u}hne}}]{2011Taming}%
  \BibitemOpen
  \bibfield  {author} {\bibinfo {author} {\bibfnamefont {B.}~\bibnamefont
  {Jungnitsch}}, \bibinfo {author} {\bibfnamefont {T.}~\bibnamefont
  {Moroder}},\ and\ \bibinfo {author} {\bibfnamefont {O.}~\bibnamefont
  {G\"{u}hne}},\ }\bibfield  {title} {\bibinfo {title} {Taming multiparticle
  entanglement},\ }\bibfield  {journal} {\bibinfo  {journal} {Physical Review
  Letters}\ }\textbf {\bibinfo {volume} {106}},\ \href
  {https://doi.org/10.1103/physrevlett.106.190502}
  {10.1103/physrevlett.106.190502} (\bibinfo {year} {2011})\BibitemShut
  {NoStop}%
\bibitem [{\citenamefont {Jungnitsch}(2011)}]{pptmixer}%
  \BibitemOpen
  \bibfield  {author} {\bibinfo {author} {\bibfnamefont {B.}~\bibnamefont
  {Jungnitsch}},\ }\href
  {https://www.mathworks.com/matlabcentral/fileexchange/30968-pptmixer-a-tool-to-detect-genuine-multipartite-entanglement}
  {\bibinfo {title} {$\mathrm{PPTM}$ixer: A tool to detect genuine multipartite
  entanglement}} (\bibinfo {year} {2011})\BibitemShut {NoStop}%
\bibitem [{\citenamefont {L{\"{o}}fberg}(2004)}]{Lofberg2004}%
  \BibitemOpen
  \bibfield  {author} {\bibinfo {author} {\bibfnamefont {J.}~\bibnamefont
  {L{\"{o}}fberg}},\ }\bibfield  {title} {\bibinfo {title} {$\mathrm{YALMIP}$:
  A toolbox for modeling and optimization in $\mathrm{MATLAB}$},\ }in\
  \href@noop {} {\emph {\bibinfo {booktitle} {In Proceedings of the CACSD
  Conference}}}\ (\bibinfo {address} {Taipei, Taiwan},\ \bibinfo {year}
  {2004})\BibitemShut {NoStop}%
\bibitem [{\citenamefont {Toh}\ \emph {et~al.}(1999)\citenamefont {Toh},
  \citenamefont {Todd},\ and\ \citenamefont {T\"{u}t\"{u}nc\"{u}}}]{Tutuncu99}%
  \BibitemOpen
  \bibfield  {author} {\bibinfo {author} {\bibfnamefont {K.~C.}\ \bibnamefont
  {Toh}}, \bibinfo {author} {\bibfnamefont {M.~J.}\ \bibnamefont {Todd}},\ and\
  \bibinfo {author} {\bibfnamefont {R.~H.}\ \bibnamefont
  {T\"{u}t\"{u}nc\"{u}}},\ }\bibfield  {title} {\bibinfo {title}
  {$\mathrm{SDPT3}$ -- a $\mathrm{M}$atlab software package for semidefinite
  programming, $\mathrm{V}$ersion 1.3},\ }\href
  {https://doi.org/10.1080/10556789908805762} {\bibfield  {journal} {\bibinfo
  {journal} {Opt. Meth. Softw.}\ }\textbf {\bibinfo {volume} {11}},\ \bibinfo
  {pages} {545} (\bibinfo {year} {1999})}\BibitemShut {NoStop}%
\bibitem [{\citenamefont {T\"{u}t\"{u}nc\"{u}}\ \emph
  {et~al.}(2003)\citenamefont {T\"{u}t\"{u}nc\"{u}}, \citenamefont {Toh},\ and\
  \citenamefont {Todd}}]{Tutuncu03}%
  \BibitemOpen
  \bibfield  {author} {\bibinfo {author} {\bibfnamefont {R.~H.}\ \bibnamefont
  {T\"{u}t\"{u}nc\"{u}}}, \bibinfo {author} {\bibfnamefont {K.~C.}\
  \bibnamefont {Toh}},\ and\ \bibinfo {author} {\bibfnamefont {M.~J.}\
  \bibnamefont {Todd}},\ }\bibfield  {title} {\bibinfo {title} {Solving
  semidefinite-quadratic-linear programs using $\mathrm{SDPT3}$},\ }\href
  {https://doi.org/10.1007/s10107-002-0347-5} {\bibfield  {journal} {\bibinfo
  {journal} {Math. Program., Ser. B}\ }\textbf {\bibinfo {volume} {95}},\
  \bibinfo {pages} {189} (\bibinfo {year} {2003})}\BibitemShut {NoStop}%
\end{thebibliography}
%

\end{document}